\documentclass[%
 aip,
 amsmath,amssymb,
preprint,%
]{revtex4-2}

\usepackage{graphicx}% Include figure files
\usepackage{dcolumn}% Align table columns on decimal point
\usepackage{bm}% bold math
\usepackage[utf8]{inputenc}
\usepackage[T1]{fontenc}
\usepackage{mathptmx}
\usepackage{etoolbox}
\usepackage[T1]{fontenc}
\usepackage[dvipsnames]{xcolor}
\usepackage[english]{babel}
\usepackage{graphicx}
\usepackage{rotating} %sidewaystable
\usepackage{float}
\usepackage{placeins}
\usepackage[colorlinks, linkcolor = blue, citecolor = blue, filecolor = blue,
urlcolor =blue]{hyperref}
\usepackage{subcaption}

\usepackage{ragged2e}
\usepackage{braket}
\usepackage{amsmath,mathtools}
\usepackage{chemformula}
\usepackage{fancyhdr}
\usepackage{bm}
\usepackage{booktabs}
\usepackage{makecell}
\usepackage{array}
\usepackage{soul}

\makeatletter
\def\@email#1#2{%
 \endgroup
 \patchcmd{\titleblock@produce}
  {\frontmatter@RRAPformat}
  {\frontmatter@RRAPformat{\produce@RRAP{*#1\href{mailto:#2}{#2}}}\frontmatter@RRAPformat}
  {}{}
}%
\makeatother
\begin{document}
\makeatletter
\long\def\@makecaption#1#2{%
  \par\small\RaggedRight
  \textbf{#1.} #2\par
}
\makeatother

%\preprint{AIP/123-QED}

\title{Excitation of Low-Frequency Modes and the Effects of Protein Dynamics on Spectral Densities of Bacteriochlorophyll Molecules}

% Force line breaks with \\
\author{Sayan Maity*}
\affiliation{School of Science, Constructor University, Campus Ring 1, 28759 Bremen, Germany}
\affiliation{Department of Physics and Astronomy and Thomas Young Centre, University College London, London WC1E 6BT, U.K.}
\email{smaity@constructor.university}

\author{Tristan A. Mauck}
\affiliation{School of Science, Constructor University, Campus Ring 1, 28759 Bremen, Germany}

\author{Ulrich Kleinekath\"ofer*}
\affiliation{School of Science, Constructor University, Campus Ring 1, 28759 Bremen, Germany}
\email{ukleinekathoefer@constructor.university}


\begin{abstract}
 In the theory of open quantum systems, spectral densities are key quantities for modeling the dynamics and spectroscopic properties of the system under investigation. In the case of light-harvesting complexes, they encode the frequency-dependent coupling of electronic excitations in pigment molecules to their environment, reflecting contributions from both intrinsic vibrational modes and the protein surrounding. In particular, the low-frequency components of the spectral densities are crucial for exciton transfer between pigment molecules. Apparently, slow internal modes of bacteriocholophyll molecules in the gas phase are less well represented by common force fields based on classical molecular dynamics (MD) simulations.  Here, we demonstrate that  Born-Oppenheimer molecular dynamics (BOMD) based on the numerically efficient density functional-based tight-binding approach can accurately recover these low-frequency features, whereas normal mode analysis captures them only partially. In contrasting approaches for determining spectral densities, the low-frequency region of the spectral densities obtained is only associated with protein fluctuations; the usage of BOMD, however, also captures the low-frequency contributions arising from slow intramolecular vibrations of the pigment molecules themselves. Notably, this behavior is consistently observed for both the flexible B800 and the more rigid B850 rings in light-harvesting 2 (LH2) complexes of purple bacteria, as well as in the Fenna-Matthews-Olson (FMO) complex of green sulfur bacteria. Interestingly, we also find that the spectral densities of the pigments in the B850 ring of LH2 are not influenced by the environment, i.e., the gaps between ground and first excited state are not changed significantly by the fluctuations of the protein environment.
\end{abstract}

\maketitle

% ----- Communication header on left -----
\pagestyle{fancy}
\fancyhf{}
%\fancyhead[R]{\textit{Communication}}
\fancyfoot[C]{\thepage}
\renewcommand{\headrulewidth}{0pt}
% Force this style also on the first page
\thispagestyle{fancy}

\section{Introduction}
Chlorophyll (Chl) and bacteriochlorophyll (BChl) molecules are the primary pigments responsible for light absorption and excitation energy transfer within the light-harvesting (LH) antenna complexes of plants and bacteria. Although experimental investigations of energy transfer processes in LH complexes using ultrafast pump-probe techniques have already been conducted over the past couple of decades \cite{cogd03,cogd06a,mirk17a}, theoretical and computational studies of antenna complexes on the molecular level have only expanded rapidly in recent years \cite{curu17a,sega19a,mait23a,jha26a}. This growth in theoretical and computational modeling has been driven by extensive three-dimensional structural characterization of antenna complexes across plants, algae, and bacteria\cite{loks21a}. Some notable examples of LH antenna complexes include the LH2 complexes of purple bacteria, the Fenna-Matthews-Olson (FMO) complex of green sulfur bacteria, and the major light-harvesting antenna complex of higher plants, LHCII, all of which were studied extensively in both theory and experiment\cite{abra09a,jank11a,kond17a}. On the experimental side, state-of-the-art ultrafast spectroscopic techniques such as two-dimensional electronic spectroscopy (2DES) and two-dimensional electronic-vibrational spectroscopy (2DEV) have specifically been tailored to investigate the intrinsic mechanisms governing the excitation energy transfer in antenna complexes\cite{lewi15a,gelz19a,arse20a,bisw22a}.   Atomic-level simulations employ either force-field-based molecular dynamics (MD) simulations or semi-empirical Born-Oppenheimer molecular dynamics (BOMD). Such simulations are usually combined with excited-state calculations using contemporary semi-empirical schemes or time-dependent density functional theory (TD-DFT) with computationally tractable functionals\cite{cign22a,mait23a}. Only in some cases, non-adiabatic simulations have been performed for LH systems, which is challenging partially because of the sheer size of the complexes \cite{sist17a,soko24a,hoff25a}. 
Within the framework of open quantum systems, the key parameters include the excitation energies, also known as site energies, the excitonic couplings between the pigments, and the so-called spectral densities of the pigment molecules  \cite{may23a}. These quantities are used to model exciton dynamics using a Frenkel Hamiltonian of the form
\begin{equation}
H = \sum_{m} E_m \ket{m}\bra{m} + \sum_{m \neq n} V_{mn} \ket{n}\bra{m}~,
\label{eq:hamil}
\end{equation}
where the diagonal elements $E_m$ represent the site energies of pigment $m$ and the off-diagonal elements $V_{mn}$ denote the inter-pigment couplings. This system part is subsequently coupled to a thermal bath with a spectral density describing the coupling.

The large size and complex electronic structure of LH complexes, but also the pigment molecules themselves require a combination of classical and quantum approaches 
to extract the relevant parameters and to determine the exciton dynamics based on atomic-level simulations. A rather straightforward but approximate scheme for the exciton dynamics is the  Numerical Integration of the Schrödinger Equation (NISE) approach \cite{jans06a,aght12a}. This scheme makes use of the full time-dependence of the Hamiltonian in Eq.~\ref{eq:hamil} and treats the coupling to the environment through fluctuations mainly in the site energies. 
Alternatively, a time-averaged Hamiltonian can be used with the fluctuations being mapped into spectral densities.  
Numerous theoretical methods have been developed to determine the exciton dynamics and spectroscopic properties using the latter approach including several variations of Redfield theory, Förster theory but also numerical exact approaches including the Hierarchy Equations of Motion (HEOM) approach \cite{tani20a}, the Hierarchy of Pure States (HOPS) \cite{gera23a} and very recent advances \cite{lore25a}. Over the last decades, a multitude of 
 reports on the excited-state dynamics of light-harvesting antenna complexes have been published \cite{aght12a,jang18a,mait23a}. However, sophisticated mixed quantum-classical techniques, such as non-adiabatic dynamics including surface-hopping methods, have only recently been applied to perform exciton dynamics on the FMO\cite{soko24a} and LH2\cite{hoff25a} complexes, where the system Hamiltonian is calculated in an “on-the-fly'' manner. However, these approaches are numerically demanding; therefore, machine learning techniques\cite{soko24a,hoff25a} or GPU acceleration\cite{sist17a} have been employed to enable efficient simulations of the real-time dynamics.

In open quantum systems, the spectral density plays a key role in describing the interaction between a system and its surrounding thermal bath. It is a frequency-dependent, temperature-independent function that characterizes the system-bath coupling and captures the environmental fluctuations that govern the exciton dynamics through the evolution of the density matrix.
It assumes linear coupling between the system and bath modes with coupling constants $c_{j\xi}$.
Denoting the mass of the bath   oscillator by $m_\xi$  and its frequency by
$\omega_\xi$, the spectral density of pigment $j$ embedded in a harmonic bath is given by
\begin{eqnarray}
\label{jw_d}
J_{j}(\omega)=\frac{\hbar}{\pi} J_{CL,j}(\omega) =\frac{1}{2} \sum_{\xi} \frac{c_{j\xi}^2}{m_\xi \omega_\xi} \delta(\omega-\omega_\xi)~,
\end{eqnarray}
where $J_{CL,j}(\omega)$ is the spectral density in the Caldeira-Legett model
that differs from the present definition by a constant factor.
%The spectral density is defined as\cite{may11a}
%\begin{equation}
%J_m(\omega) = \frac{1}{2} \sum_{k} c_{m,k}^2 \, \delta(\omega - \omega_k)~,
%\end{equation}
%where \(c_{m,k}\) are dimensionless coupling parameters associated with the vibrational frequencies \(\omega_k\).
Here, one uses the fact that the interaction between the system and the environment in systems such as light-harvesting complexes is diluted over a large number of degrees of freedom.  In this case, an effective harmonic bath produces a system dynamics which equals that of the actual bath, potentially including anharmonic modes \cite{makr24a,makr24b}. 
Recently, it has, however, been shown that  anharmonic couplings between
pigment vibrational modes cannot be captured by the harmonic oscillator bath \cite{cho25a}, a topic which needs further investigation.

In experiments, spectral densities can be reconstructed based on mode-resolved Huang-Rhys (HR) factors extracted from fluorescence line-narrowing or spectral hole-burning measurements\cite{kell13a,piep18a}. In a computational framework, they can be determined as the half-sided Fourier transform of the autocorrelation function of site energy fluctuations. This is expressed as
\begin{equation}
J_j(\omega) = \frac{\beta \omega}{\pi} \int_0^{\infty} dt \, C_j(t) \cos(\omega t)~,
\label{eq:spd_intro}
\end{equation}
where \(\beta\) denotes the inverse temperature and \(C_j(t)\)  the autocorrelation function of the site energy fluctuations for the pigment molecule \(m\).  It is defined by
\begin{equation}
C_j(t_l) = \frac{1}{N-l} \sum_{k=1}^{N-l} \Delta E_j(t_l + t_k) \, \Delta E_j(t_k)~.
\label{eq:acf_intro}
\end{equation}
Here, \(\Delta E_j = E_j - \langle E_j \rangle\) represents the deviation of the site energy \(E_j\) from its average value, and \(N\) is the number of snapshots along the trajectory. 

Traditionally, classical molecular dynamics (MD) simulations, followed by excitation energy calculations within a quantum mechanics/molecular mechanics (QM/MM) framework, either using semi-empirical methods or density functional theory, have been employed to extract site energy fluctuations and, consequently, the spectral density\cite{olbr11b,zueh19a,zueh19b}. However, this procedure suffers from
at least two issues: 
the so-called “geometry-mismatch'' problem and a poor representation of internal vibrational modes of the pigments by common force fields. The first issue describes potentially poor results when quantum mechanical calculations for excited states are performed on ground state geometries generated based on classical force fields \cite{juri15a,padu17a}. The second issue often introduces artifacts in the high-frequency intramolecular vibrational modes of pigment molecules in the resulting spectral density. To address this issue in ground-state dynamics, numerically inexpensive semi-empirical methods such as PM6 \cite{rosn15a}, or more computationally demanding DFT approaches employing hybrid functionals like B3LYP \cite{blau18a} have been used within a QM/MM framework. However, these methods are limited by either insufficient accuracy or high computational cost. Moreover, an interpolated potential energy surface has been constructed at the DFT level for both the  S$_0$ ground  and the  S$_1$ excited states and combined with an MM environment to enable long-time molecular dynamics simulations\cite{kim16a, kim18a}. However, constructing such  interpolated energy surfaces is technically demanding, as it requires numerous reference DFT and TD-DFT calculations to adequately sample the relevant conformational space, particularly for a large pigment molecule such as BChl. Recently, we have addressed this issue by performing numerically efficient semi-empirical density functional-based tight-binding (DFTB) force evaluations to propagate ground-state MD within a QM/MM framework. This approach resolves both mentioned issues to a large degree. The calculation of the ground state dynamics is
followed by excitation energy calculations using the time-dependent, long-range corrected DFTB (TD-LC-DFTB), also within a QM/MM fashion \cite{mait20a,mait21a,mait21b}. This strategy overcomes the geometry-mismatch problem and accurately reproduces the high-frequency intramolecular vibrational peaks due to fast oscillations involving C=C, C=O, and C=N double bonds of the pigment molecules observed in experiments\cite{mait20a,mait21a}. In previous studies, we have successfully applied this approach to bacterial, plant, and algae systems \cite{mait23a}. The scheme is able to capture complex features in structured spectral densities, including both high-frequency intramolecular vibrational modes (underdamped components) and low-frequency intermolecular motions arising from electrostatic coupling to environmental fluctuations (overdamped components).
However, it is known that the reorganization energies of pigment molecules are somewhat overestimated in this scheme due to the choice of the semi-empirical DFTB method for both ground and excited states\cite{mait21a}. 

Spectral densities can also be determined using several alternative approaches. One such approach is based on the charge density coupling (CDC) method \cite{adol08a}. In this framework, the spectral density is obtained from fluctuations of the excitation energy arising from electrostatic interactions between the pigment and its environment. Intramolecular vibrational contributions are neglected, so that the resulting spectral density describes only environmentally induced (electrostatic) energy-gap fluctuations. This CDC scheme has been applied in a number of studies to extract excitation energies of BChl or Chl pigment molecules within a protein environment, e.g., in Refs.~\cite{jing12a,rive13a,reng13b,vegt15a,hsie19a,mika25a}.  
Subsequently, different theoretical schemes have been developed to extract intramolecular vibrational peaks using the vertical gradient (VG) approximation, the adiabatic Hessian (AH) model or excited state forces applied to individual normal modes, where the gradient of the excited-state potential energy surface is evaluated at the ground-state equilibrium geometry\cite{lee16a,lee16c,padu17a,cign22a}. 
%In further studies, the spectral density for the low-frequency part of the spectral density is supplemented by a high-frequency part from a normal mode analysis. 
These approximations can reproduce intramolecular vibrational peaks with reasonable accuracy, provided that the chosen DFT functional yields reliable normal modes for the pigment molecules. Often, the intramolecular contributions, treated within the VG approximation, are combined with the CDC-based description of low-frequency intermolecular modes to construct the full spectral density\cite{lee16a,lee16c}. Alternatively, Renger and co-workers proposed a scheme in which the intermolecular contributions to the spectral density are obtained from a normal-mode analysis of the pigment-protein complex, while the intramolecular pigment vibrations are incorporated separately using experimental data\cite{reng12b,reng13a,reng13b,klin20a}. What is missing in all of these descriptions are possible low-frequency intra-molecular vibrations in the typical frequency range from a few to several tens of cm$^{-1}$.  A key problem of these combined approaches is the assumption that inter- and intra-molecular peaks occur in different frequency ranges. This causes them to miss possible low-frequency intra-molecular vibrations in the typical frequency range from a few to several tens or even hundreds of cm$^{-1}$.  Even for apparently rigid molecules such as bacteriochlorophyll and chlorophyll pigments containing porphyrin rings, these frequencies do not correspond to the simple harmonic normal modes centered at a single equilibrium geometry but are instead associated with slow nuclear motions such as large-amplitude torsions, ring puckering, bending, and other collective intramolecular deformation modes\cite{jent97a,shel98a,jent98a}. These modes include doming, ruffling, saddling, and waving of the porphyrin (similar to the chlorin ring in the case of chlorophyll), which correspond to soft normal coordinates of the macrocycle and represent energetically favored pathways for nonplanar deformation. These low-frequency modes cannot be captured if the low-frequency region is left solely to the CDC or similar methods.   The spectral density, especially its low-frequency region due to the energetically close exciton energies, is a key ingredient in modeling excitation dynamics and the optical properties of the LH2 complex. Because of their strong anharmonic character, low-frequency slow deformation modes of pigment molecules within the protein environment are difficult to detect by existing methods based either on normal mode analysis of the protein conformational dynamics, which relies on the harmonic approximation\cite{reng12b,reng13a}, or on-site energy fluctuation approaches that consider only intermolecular interactions arising from electrostatic coupling to protein conformational dynamics \cite{jing12a,rive13a,lee16a,lee16c}. However, explicitly accounting for excitation-energy fluctuations of pigment molecules within the protein environment enables one to capture the full range of harmonic contributions, including high-frequency intramolecular vibrations and low-frequency, slow pigment deformation modes coupled to protein conformational dynamics. This is discussed in detail below for the LH2 and  FMO complexes.

In this study, we focus on the low-frequency region of the spectral density of the apparently rigid BChl pigment molecule. These low-frequency modes correspond to collective soft motions of the system rather than localized high-energy vibrations and are often believed to originate primarily from the surrounding bath, including conformational dynamics of proteins and solvents \cite{reng12b,klin20a}. However, in this study, we show that, in addition to the conformational dynamics of proteins and solvents, slow intramolecular vibrational modes of the porphyrin ring of BChl molecules, which are related to distortion, puckering, and in-plane and out-of-plane motions\cite{jent97a,shel98a,jent98a},  contribute to this frequency region as well. As will be shown below, these slow vibrational features can be observed when the ground state dynamics are determined through BOMD simulations employing the semi-empirical DFTB3 level of theory. For example, the corresponding peaks are prominently observed for the BChl molecule simulated in the gas phase, at least up to 225~cm$^{-1}$. Furthermore, we performed a normal mode analysis (NMA) of the BChl molecule in its ground and first excited states in the gas phase using DFT. Surprisingly, the NMA also shows a non-zero spectral density based on the calculated Huang-Rhys factors, which represent a dimensionless constant describing the coupling between electronic excitation and vibrational modes. Several modes appear in the low-frequency region (100-225~cm$^{-1}$) in the NMA, partially capturing the slow low-frequency modes compared to the DFTB-based simulations. 

Furthermore, the low-frequency contributions have been analyzed in detail for  BChl pigments embedded in the LH2 complex, specifically in the B800 and B850 rings of the LH2 complex of the purple bacterium \textit{Rhodospirillum molischianum}, as well as for a pigment in the Fenna-Matthews-Olson (FMO) complex of the green sulfur bacterium \textit{Chlorobaculum tepidum}. For the LH2 complex, the low-frequency spectral density of the B800 ring is more intense in DFTB-based QM/MM MD than in classical MD, reflecting underestimated intramolecular pigment contributions in the latter, while protein conformational dynamics are captured in both methods. In contrast, in the more rigid B850 ring of the same complex, the low-frequency peaks are less pronounced in classical MD than the B800 ring but become significantly intensed in DFTB/MM MD compared to the classical MD due to slow vibrational motions of the BChl pigment captured better in DFTB-based method similar in case of the B800 ring. Most importantly, in the case of the B850 ring, the  conformational dynamics of the surrounding environment has a negligible effect.  Rather, the low-frequency region of the spectral density in both classical MD and DFTB/MM MD simulations for this ring is mainly contributed to by the slow intramolecular modes of the pigments. Nevertheless, both classical MD and DFTB/MM MD can capture the low-frequency intramolecular modes of bacteriochlorophyll within the protein environment, similarly to gas-phase calculations. As a second example, in the case of the FMO complex, the low-frequency modes are dominated primarily by the protein environment in classical MD simulations, whereas in DFTB-based QM/MM MD, they include contributions from both the protein and the pigment, similar to the behavior observed for the B800 ring in the LH2 complex.

\section{\label{sec:method} Computational Methods}
As a starting point for characterizing the low-frequency modes of BChl, we truncated the phytyl tail at the CP1-CP2 bond, capped it with a hydrogen atom, and optimized the resulting structure at the B3LYP/def2-TZVP level using the ORCA 6 quantum chemistry code\cite{nees25a}. The optimized structure was then used for gas-phase BOMD simulations employing the semiempirical, numerically efficient 3rd-order corrected DFTB method\cite{gaus11a} with the frequency-corrected 3OB-f parameter set\cite{gaus13a}. The DFTB simulations were performed using the DFTB+ code \cite{hour20a}. Ten trajectories of 200~ps each were generated, with initial velocities assigned according to a Maxwell-Boltzmann distribution at 300~K. To avoid perturbing the dynamics, no thermostat was applied, resulting in simulations in the NVE ensemble. Snapshots were saved every 5~fs and subsequently used as the basis for excitation energy calculations. To this end, we employed the semi-empirical TD-LC-DFTB method with the OB2 parameter set as implemented in the DFTB+ code to compute the 10 lowest excited states\cite{kran17a}, from which the Q${_y}$ excitation energy corresponding to the S${_1}$ transition was extracted along the trajectory. The TD-LC-DFTB method has been extensively benchmarked against other long-range-corrected DFT functionals (CAM-B3LYP and $\omega$B97X) for BChl\cite{bold20a} and Chl \cite{mait24a} molecules, demonstrating a significantly lower computational cost with only a minimal loss of accuracy. Subsequently, the autocorrelation functions and the corresponding spectral densities were computed from the ten time series of site energies, as shown in Eqs.~\ref{eq:spd_intro} and \ref{eq:acf_intro}. The spectral densities obtained from the independent trajectories were averaged to yield the final spectral density. This procedure helps to reduce the noise inherent in spectral densities derived from gas-phase simulations. As a supplementary analysis, we performed a more computationally demanding 10~ps BOMD simulation using the DFT [BLYP/6-31G(d,p)] level of theory. The simulation was also carried out in the NVE ensemble, with snapshots saved every 1~fs, following the same procedure to extract the spectral density as shown in the SI. In all BOMD simulations, a 0.5~fs integration time step was used, and no constraints were applied.

To perform normal mode analysis, the Hessians of the ground and S$_1$ excited states were extracted at the CAM-B3LYP/def2-TZVP level of theory following geometry optimization. This functional was selected due to its documented good performance for the BChl molecule\cite{raet11a}. Moreover, TD-LC-DFTB has been benchmarked by us for the BChl molecule against this functional\cite{bold20a}, demonstrating very good accuracy and thereby justifying its use for the analysis of NMA. Mode-resolved Huang-Rhys factors were then derived from the displacement between the ground- and excited-state equilibrium geometries expressed in the normal-mode basis. All calculations were performed using ORCA 6 quantum chemistry code\cite{nees25a}, with the electronic dependence on nuclear modes evaluated using the Excited State Dynamics (ESD) module\cite{souz18a}. The adiabatic Hessian (AH) model was employed, and Duschinsky mixing between ground- and excited-state modes was included.

To examine the effect of conformational dynamics of the protein on the low-frequency region in the spectral density, we used an initial structure of the LH2 complex from \textit{Rh.~molischianum} (PDB: 1LGH) employed in our previous study \cite{olbr10a}, and selected one BChl pigment each from the B800 and the B850 ring as systems of interest. Furthermore, we truncated the BChl molecule at the CP1-CP2 bond and capped the fragment with a hydrogen atom. The truncated BChl pigment was treated as the QM region, whereas the remaining phytyl tail, along with the protein, lipids, solvent, and all other BChl pigments, were treated as the MM region. To obtain the spectral density of this pigment, we equilibrated the initial structure from our previous study \cite{olbr10a} over a total NPT simulation time of 20~ns with position restraints on lipid head groups, protein side chains, protein backbones and pigment atoms lifted at every 5~ns. This was followed by another 20~ns of simulation without any restraints. The structure was modeled using the CHARMM27 force field together with a CHARMM-compatible force field for the pigments \cite{damj02a}. The Langevin thermostat as well as the Nose-Hoover Langevin barostat were employed. We subsequently performed the 50~ps QM/MM molecular dynamics simulations from the equilibrated structure. The simulations were carried out in the NPT ensemble using the semi-empirical DFTB method with the 3OB-f parameter set. All simulations were carried out with an in-house interface of DFTB+  to the NAMD3 simulation engine, based on the custom implementation of the QM/MM interface in NAMD \cite{melo18a}. In addition, a 50~ps classical MD simulation of the LH2 complex was performed using NAMD3. All DFTB/MM MD and classical MD simulations of the LH2 complex were performed in the NPT ensemble with a 0.5~fs integration time step and without any constraints. It should be further noted that in the QM/MM MD simulations based on DFTB, the pigment molecule of interest is treated as the QM region and propagated using DFTB forces, while the surrounding environment is described by classical force fields. In contrast, the classical MD simulations employ pure-MM MD, where both the pigment and the environment are propagated using classical force fields. Snapshots from all simulations were saved every 1~fs and subsequently used as bases of the excited-state calculations using the TD-LC-DFTB approach with the OB2 parameter set, as implemented in \textsc{DFTB+} within the QM/MM framework. Similar to the gas-phase calculations, the lowest singlet excited state S$_1$, corresponding to the Q$_y$ transition, was extracted along each trajectory. Again, the resulting site energy trajectories were used to compute the autocorrelation function and, ultimately, the spectral density with and without environmental point charges. In addition, DFTB/MM MD trajectories of the FMO complex from the green sulfur bacterium \textit{C. tepidum} were taken from our previous study where the system was equilibrated based on CHARMM36m force field\cite{mait20a}. We considered the so-called BChl 3 from the FMO complex and extracted the spectral density both with and without the protein environment based on TD-LC-DFTB calculations along classical MD and DFTB-based QM/MM MD trajectories. These trajectories were generated using the GROMACS simulation engine\cite{abra15a} together with a QM/MM interface to the DFTB+ program\cite{kuba15b}.

\section{\label{sec:results_disscussion}Results and Discussion}
\begin{figure} [ht!]
\centering \includegraphics[width=0.85\textwidth]{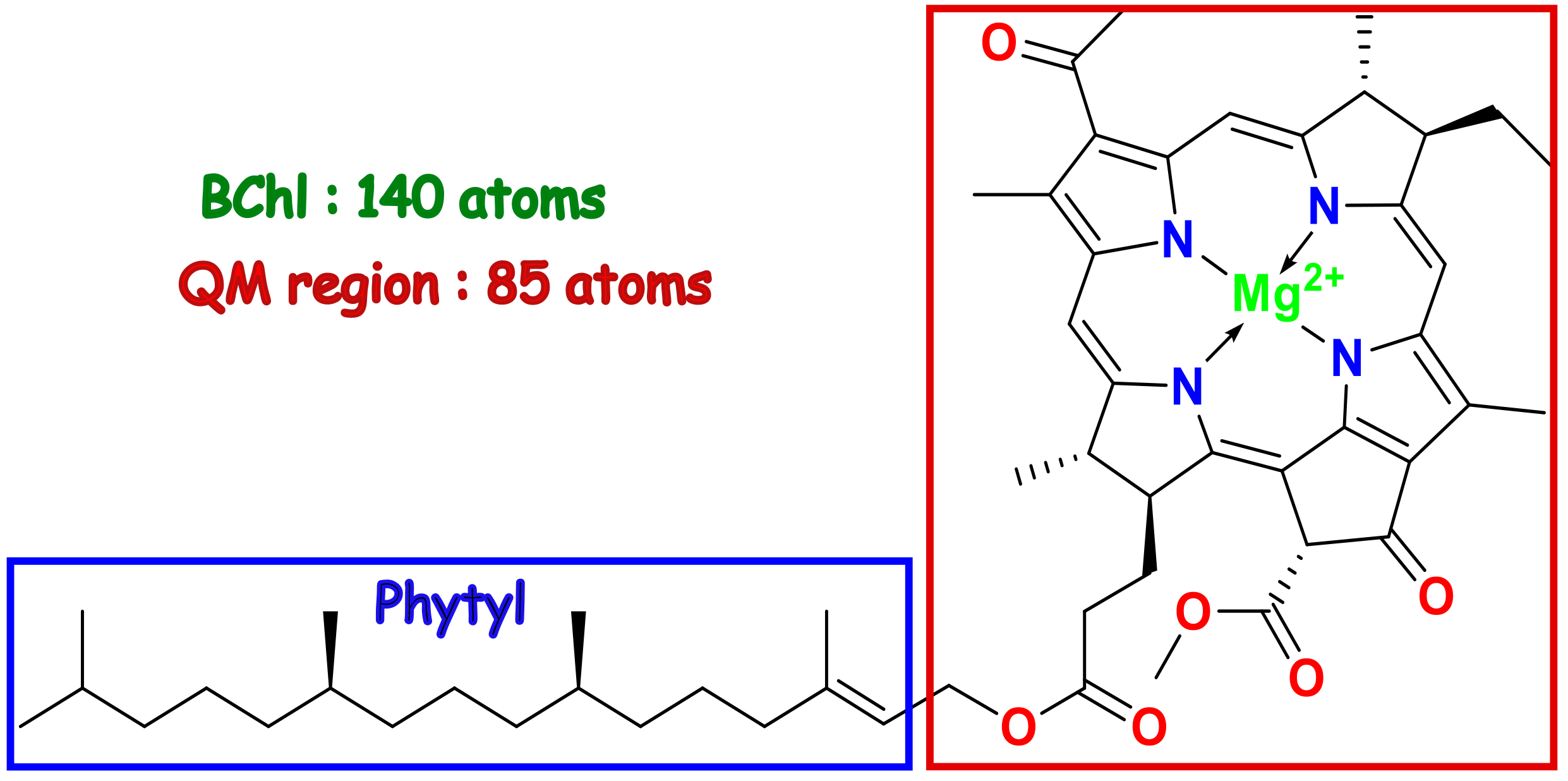}
\caption{\label{fig:bcl_struct} Two-dimensional chemical structure of the BChl pigment (which is actually a BChl a) present in the LH2 and FMO complexes. Of the total 140 atoms in the pigment, 85 are considered to be part of the QM region (indicated by the red box) for both the gas phase and the QM/MM calculations. The phytyl tail (highlighted by the blue box) is neglected in the gas-phase calculations, whereas in the QM/MM simulations it is included as part of the MM region.}
\end{figure}

The description of the results is divided into three parts. First, we compare spectral densities from BOMD simulations based on different QM methods and from classical MD in gas-phase calculations of BChl pigments. In the next two steps, we followed the same procedure for  BChl pigments within two protein systems: the LH2 complex of the purple bacterium \textit{Rs.\ molischianum} and the FMO complex of the green sulfur bacterium  \textit{C. tepidum}.
In all these simulations, the tails of the BChl~a pigments were treated as point charges. The splitting into QM and MM regions is shown in Fig.~\ref{fig:bcl_struct}.

\subsection{Spectral Density of BChl Pigments in the Gas Phase}
\begin{figure}[ht!]
    \centering
    \begin{subfigure}[t]{0.49\textwidth}
        \centering
        \includegraphics[width=\textwidth]{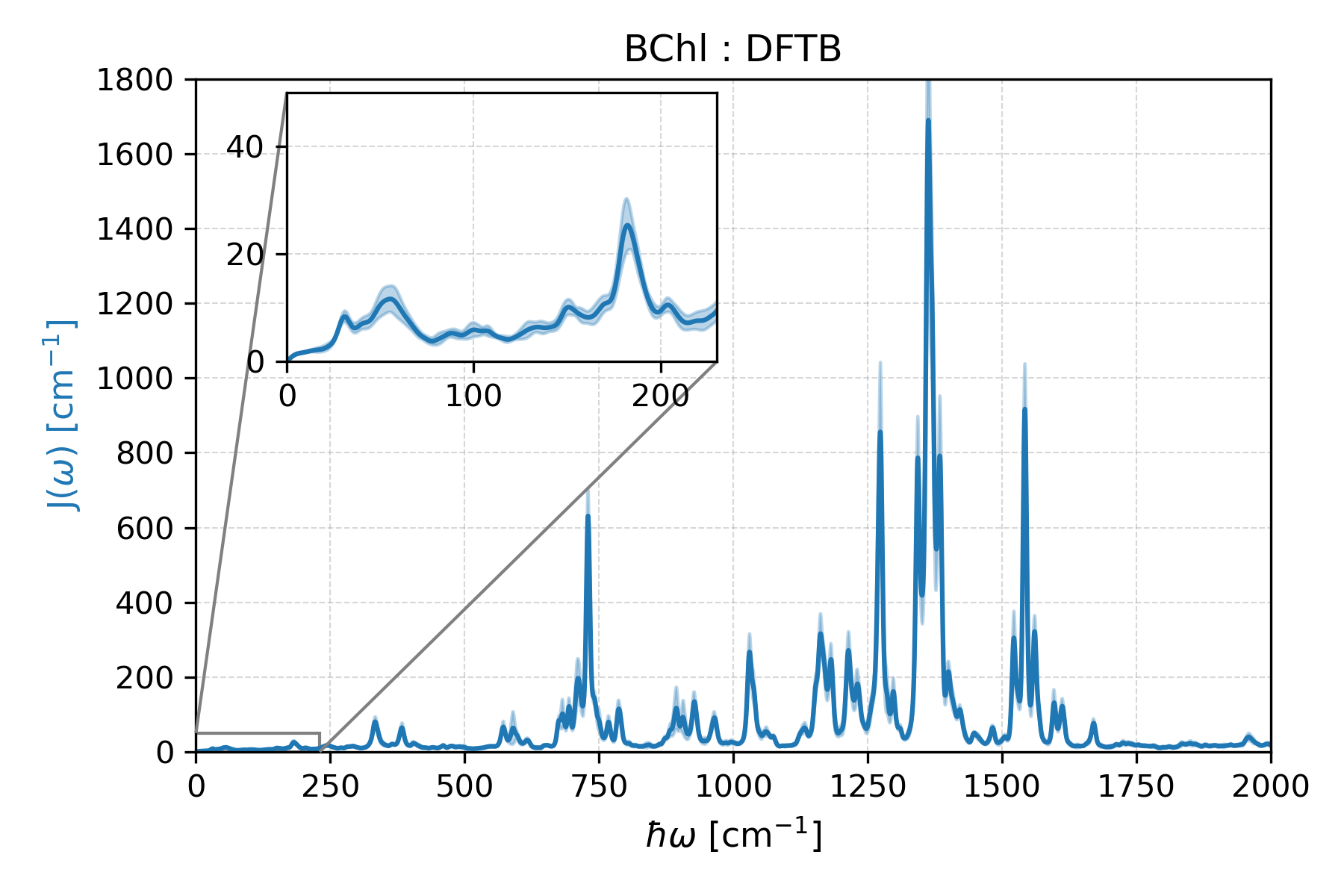}
        \caption{}
    \end{subfigure}\hfill
    \begin{subfigure}[t]{0.49\textwidth}
        \centering
        \includegraphics[width=\textwidth]{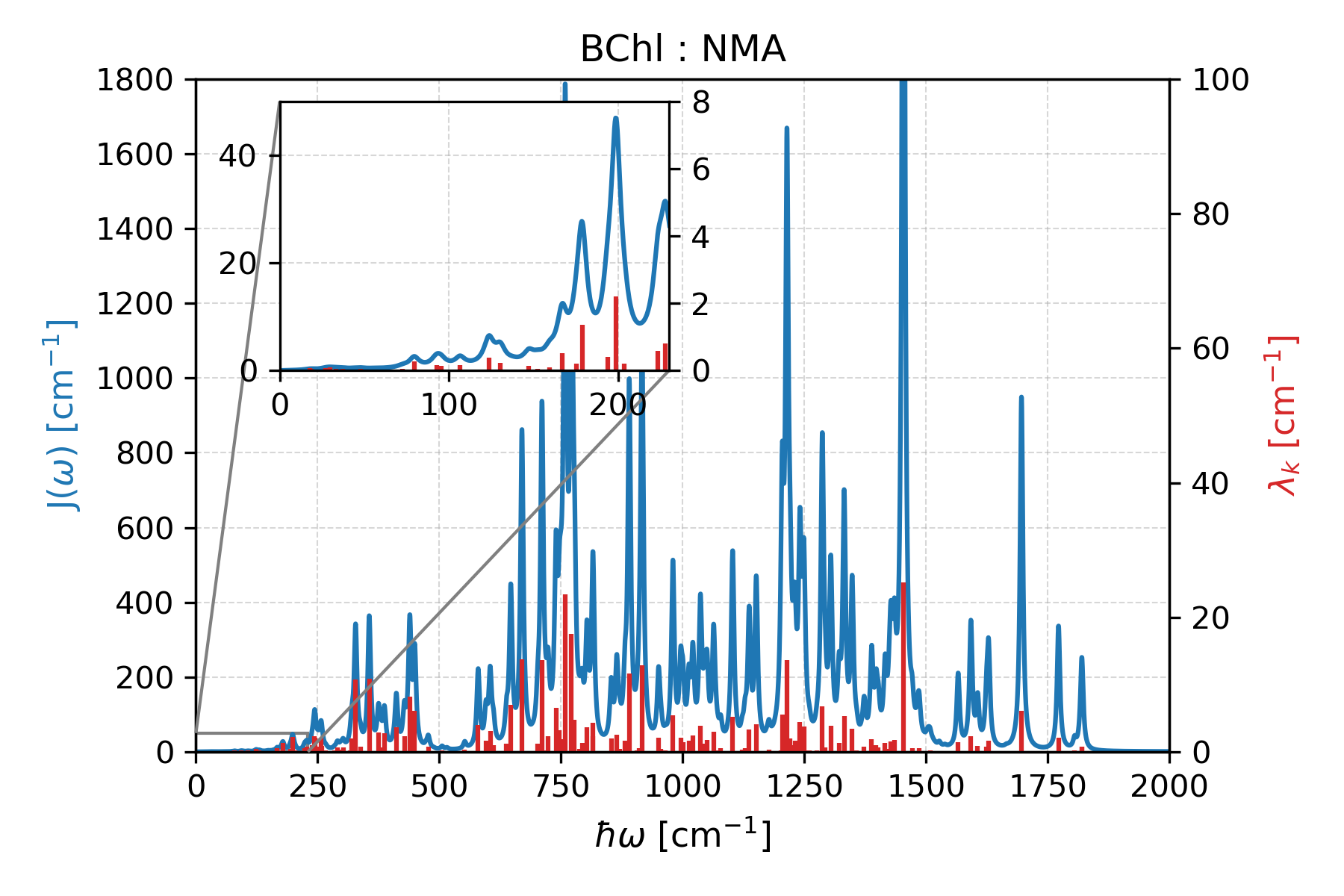}
        \caption{}
    \end{subfigure}

\caption{\label{fig:spd_dft} (A) Gas phase spectral density of the BChl molecule derived from the average of 10 NVE trajectories started at randomized starting velocities simulated at the DFTB/3OB-f level. The shaded region indicates the standard deviation from the mean. (B) Gas-phase spectral density of the BChl molecule obtained from normal mode analysis of the ground and first excited states at the CAM-B3LYP/def2-TZVP level of theory using the adiabatic Hessian model. The spectral density (blue) was constructed from the HR factors using a Lorentzian broadening function as explained in the SI. The mode-resolved reorganization energies are also shown as red bars for comparison.}
\label{fig:dftb_and_normal_modes}
\end{figure}

The spectral density based on the semi-empirical DFTB simulations for the ground state dynamics and TD-LC-DFTB for the excited state calculations in the gas phase is shown in Fig.~\ref{fig:dftb_and_normal_modes}A. Fig.~\ref{fig:dftb_and_normal_modes}B depicts the spectral density constructed from the normal mode-resolved HR factors obtained from the adiabatic Hessian analysis at the DFT (CAM-B3LYP/def2-TZVP) level of theory for both ground and excited states. This functional was selected because it provided the best performance among the tested functionals and has also been documented to perform well for NMA\cite{raet11a}. The spectral density was broadened using a Lorentzian function derived from the HR factors, as described in Eq.~1 of the SI. Furthermore, for clarity, the mode-resolved reorganization energies (frequency $\times$ HR) are also presented alongside the spectral density. The corresponding HR factors for the spectral densities derived from the DFTB simulations and NMA are listed in Tables~S1-S3 of the SI.

As can be seen from the insets, both methods find several  low-frequency peaks corresponding to slow vibrational modes of the rather rigid BChl pigment observed in X-ray experiments on porphyrin rings \cite{jent97a,shel98a,jent98a}.  The spectral density derived from the ensemble of DFTB simulations shows prominent modes at $\omega_k = 32\:\text{cm}^{-1}$, $\omega_k = 55\:\text{cm}^{-1}$, and $\omega_k = 183\:\text{cm}^{-1}$. The normal mode derived spectral density shows non-negligble intensity around $\omega_k \approx 100\:\text{cm}^{-1}$ with several larger peaks grouped around $\omega_k \approx 200\:\text{cm}^{-1}$. Thus, as reported earlier \cite{raet11a}, the NMA seems to only partially capture the low-frequency modes. This is likely due to the anharmonic character of these distortion motions, which is not fully described within the harmonic approximation used in NMA. Spectral densities all rely on the assumption of harmonic bath operators, but in the present scheme, this assumption is only applied in the last step of the analysis rather than from the beginning. It should be noted that the difference in intensity below $\omega_k \approx 100\:\text{cm}^{-1}$ between the DFTB and NMA results leads to significant differences in the Huang-Rhys factors in this range. This is mainly caused by the inverse dependence of the HR factor on its mode frequency. 

While the overall shapes of the spectral densities are similar, the peak intensities obtained from the NMA are larger, and a greater number of peaks are observed. This arises because NMA, by construction, evaluates the effect of the excited-state displacement on all normal modes of the system. In contrast, the DFTB dynamics-based approach samples only those modes that are initially excited or become populated through energy transfer during the relatively short simulation time. Consequently, the resulting spectral density depends on the extent of intermode coupling and the generation of a sufficiently large ensemble of independent trajectories. The spectral densities obtained from the DFTB dynamics show better agreement with experiment (see SI, Fig.~S2). However, this likely reflects the use of the 3OB-f parameter set, which was specifically optimized for such applications, as reported previously for LH complexes\cite{mait23a}. In contrast, hybrid functionals are known to shift high-frequency vibrational modes,\cite{gaus13a}, which is also observed in the present results based on CAM-B3LYP calculation.

This sampling issue seems to be more severe for BOMD simulations, as we have already observed previously in simulations of LH antenna complexes \cite{mait20a,mait21a,mait21b}. The purpose of this comparison, however, was not to compare the magnitude of the spectral density between BOMD and NMA. The point which we want to make here is that already an individual BChl molecule in the gas phase, i.e., without any protein environment, has a spectral density that is clearly non-vanishing in the low-frequency region. This is a point which will be further discussed below in the context of spectral densities, which are assuming that the low-frequency part results from the protein environment only. 

%We further extracted the spectral densities from classical MD trajectories, which are presented in the SI. As can be seen, the high-frequency peaks in the spectral densities obtained from classical MD simulations are shifted due to the so-called geometry mismatch and a poor representation of the internal vibrations in the force field artifacts discussed earlier. Nevertheless, in the low-frequency region, we still observe non-zero contributions to the spectral density, further suggesting the existence of these slow modes. However, despite the issue with the vibrational frequencies, these peaks are less intense than those obtained from the DFTB-based trajectories, likely due to the parametrization procedure that imposes constraints to keep the Mg-porphyrin moiety in the BChl pigment planar~\cite{damj02a}.
Furthermore, we extracted the spectral density from a 10~ps DFT-based trajectory computed with the BLYP functional and the 6-31G(d,p) basis set in the NVE ensemble; the results are presented in the SI (Fig.~S3). This functional is known to describe C=C, C=O, and C=N bond vibrations in organic molecules reasonably well\cite{gaus13a}. The overall peak positions are consistent with those obtained from the DFTB/3OB-f simulations. The resulting spectral density exhibits features in both the high- and low-frequency regions similar to those obtained from the methods discussed above, but certainly has sampling issues.

\subsection{Spectral Density of BChl Pigments in the LH2 Complex}

\begin{figure} [ht!]
\centering \includegraphics[width=0.60\textwidth]{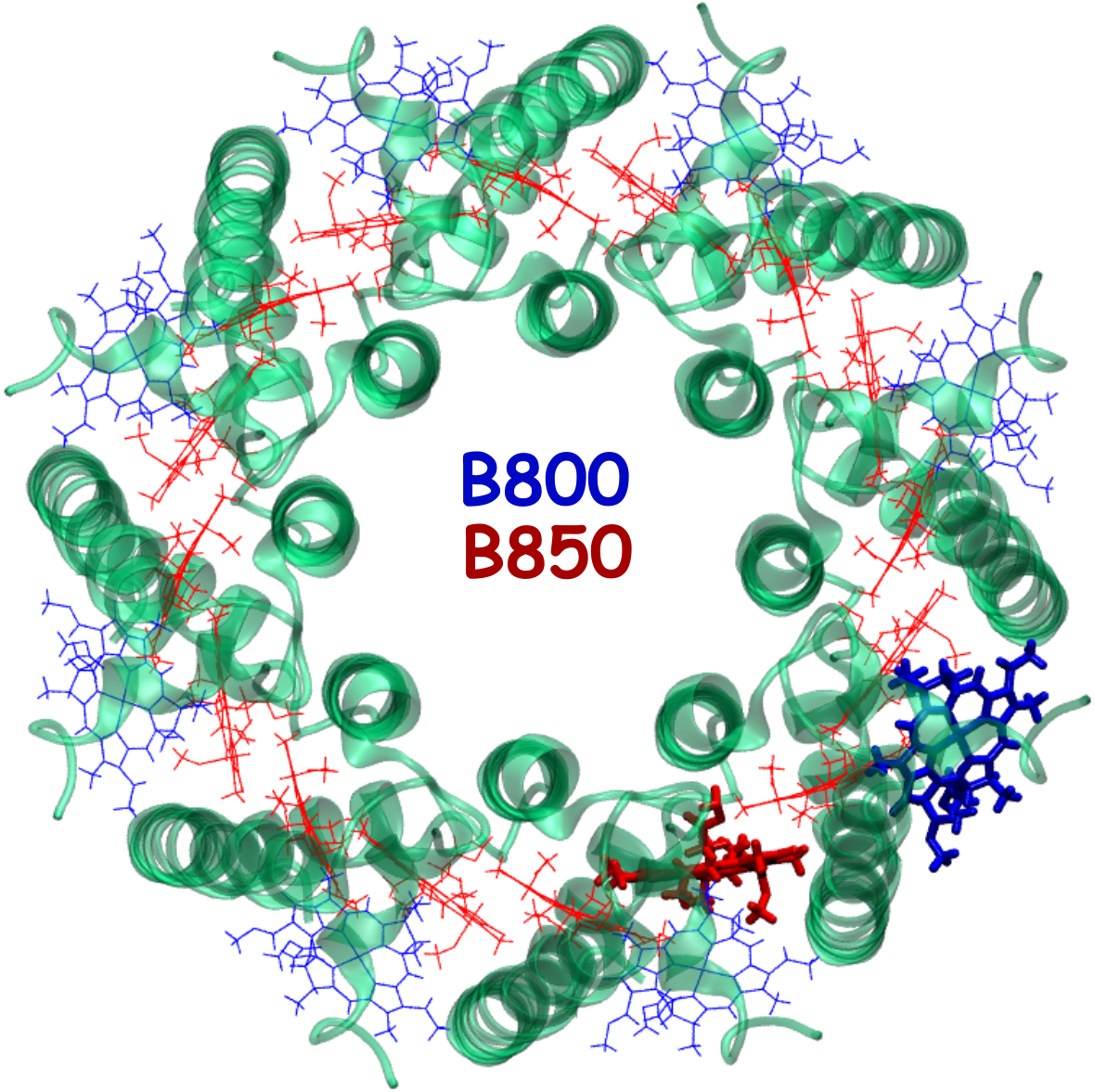}
\caption{\label{fig:lh2_sys} LH2 complex from  the purple bacterium \textit{Rs.\ molischianum} (PDB code: 1LGH), showing the B800 ring (blue, 8 pigments) and the B850 ring (red, 16 pigments). One pigment from each ring is highlighted and treated as the QM region in this study.}
\end{figure}

As a next step, we studied the excitation of the low-frequency modes of BChl pigments within antenna complexes. Specifically, the first system considered in this study is the LH2 complex of the purple bacterium \textit{Rs.\ molischanium}. This LH2 complex with eightfold symmetry contains 24 BChl molecules arranged in the B850 and B800 rings. The B800 and B850 rings are classified by their absorption wavelengths at around 800~nm and 850~nm, respectively.
The B850 ring contains 16~$\alpha\beta$ BChl pigments, while the B800 ring contains 8 BChl molecules. The LH2 complex has been extensively investigated both theoretically and experimentally, and its intrinsic dynamics and optical properties have been reported in numerous studies~\cite{hare12a,pani10a,olbr10a,olbr10b,cupe16b,cupe18b,mirk17a,cupe20b,hoff25a}.

\begin{figure*}[ht!]
    \centering

    % Left panel
    \begin{subfigure}[b]{0.50\textwidth}
        \centering
        \includegraphics[width=\textwidth]{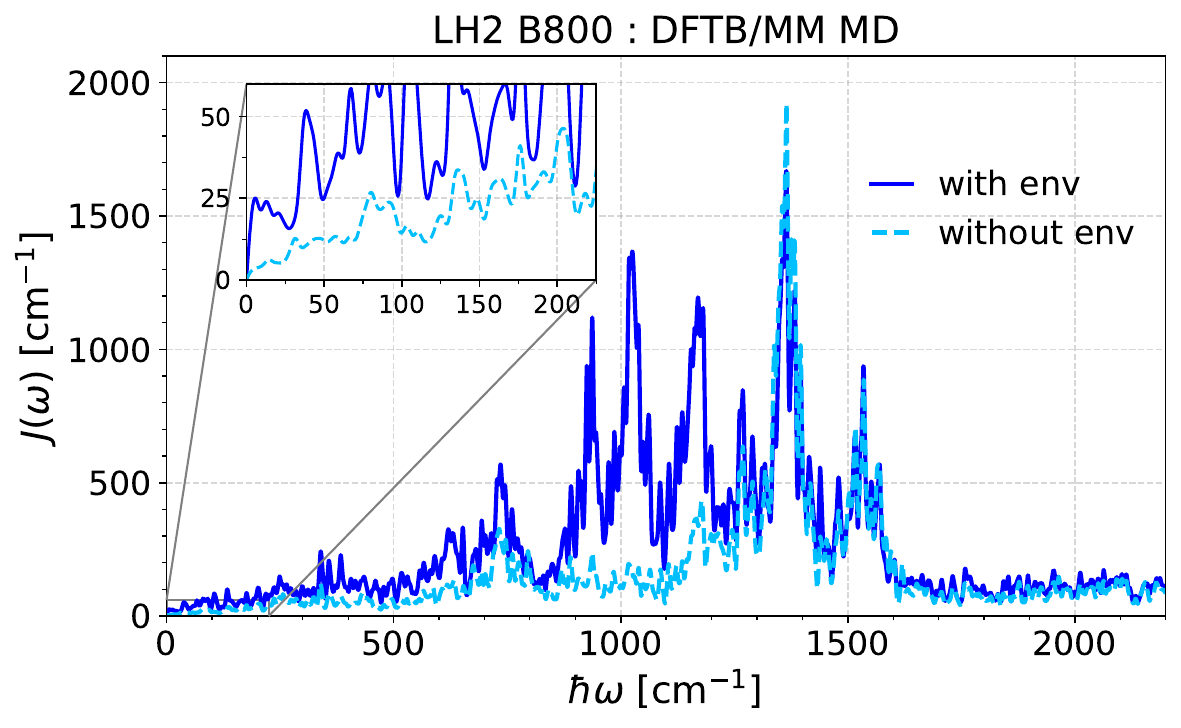}
        \caption{}
        \label{fig:B800_dftb}
    \end{subfigure}
    \hfill
    % Right panel
    \begin{subfigure}[b]{0.49\textwidth}
        \centering
        \includegraphics[width=\textwidth]{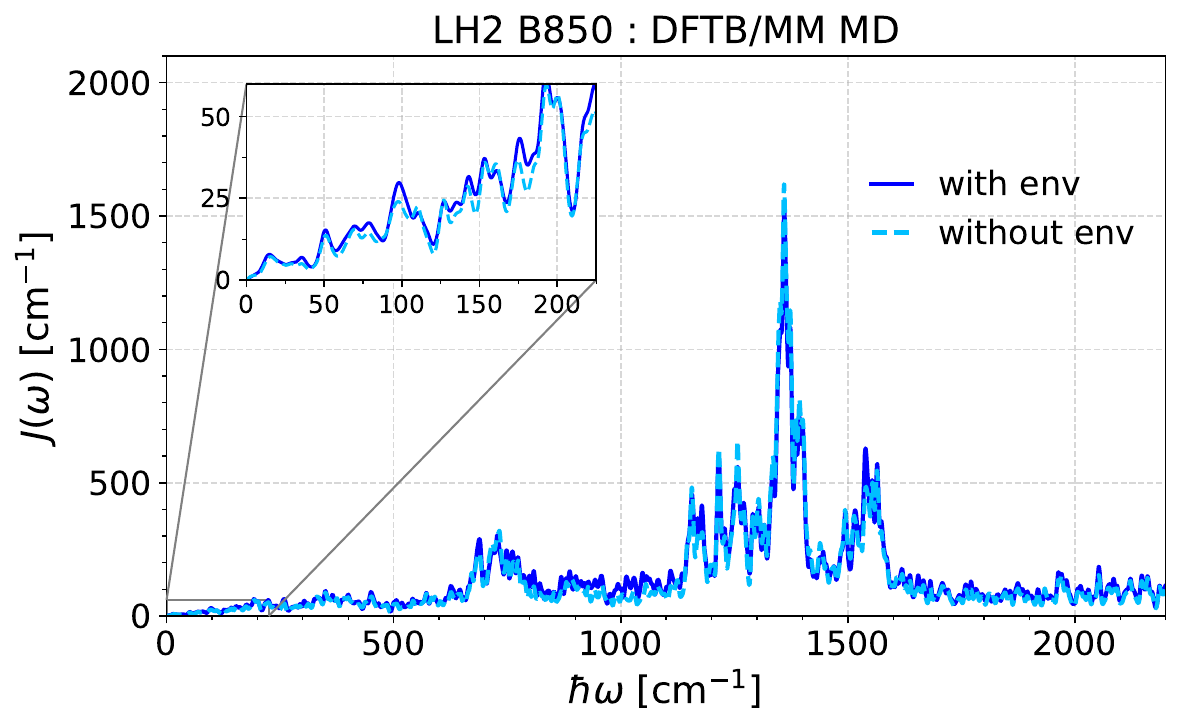}
        \caption{}
        \label{fig:B850_dftb}
    \end{subfigure}

% Left panel
    \begin{subfigure}[b]{0.50\textwidth}
        \centering
        \includegraphics[width=\textwidth]{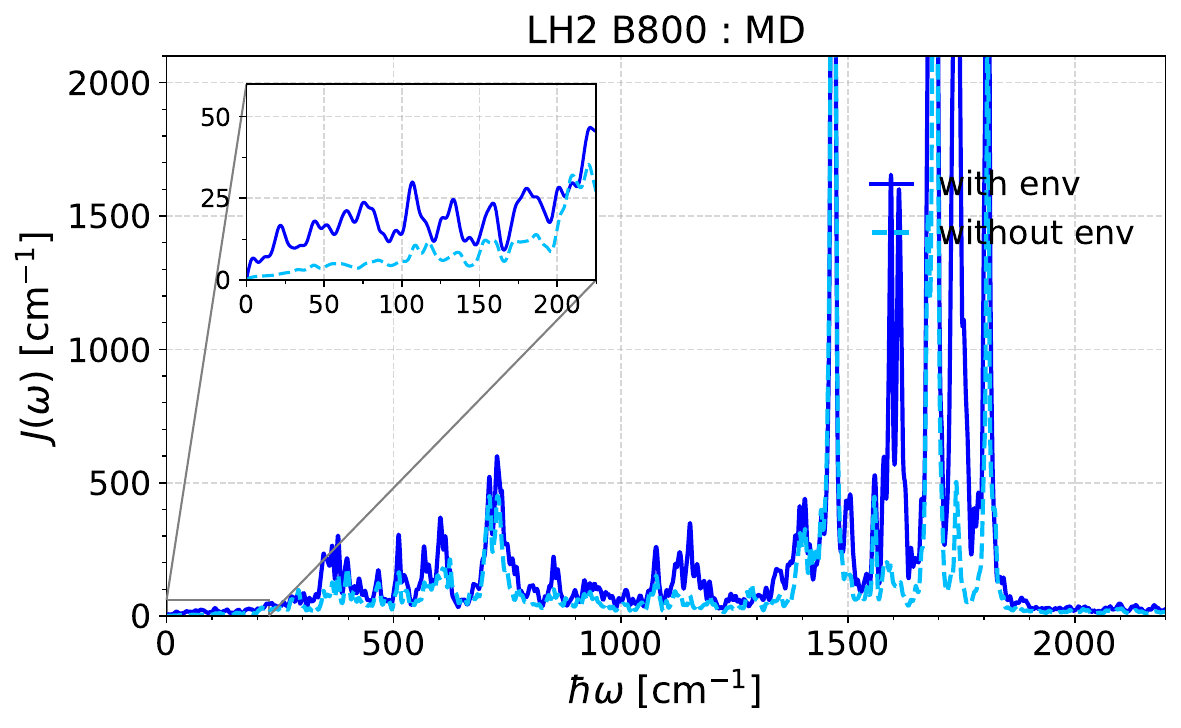}
        \caption{}
        \label{fig:B800_md}
    \end{subfigure}
    \hfill
    % Right panel
    \begin{subfigure}[b]{0.49\textwidth}
        \centering
        \includegraphics[width=\textwidth]{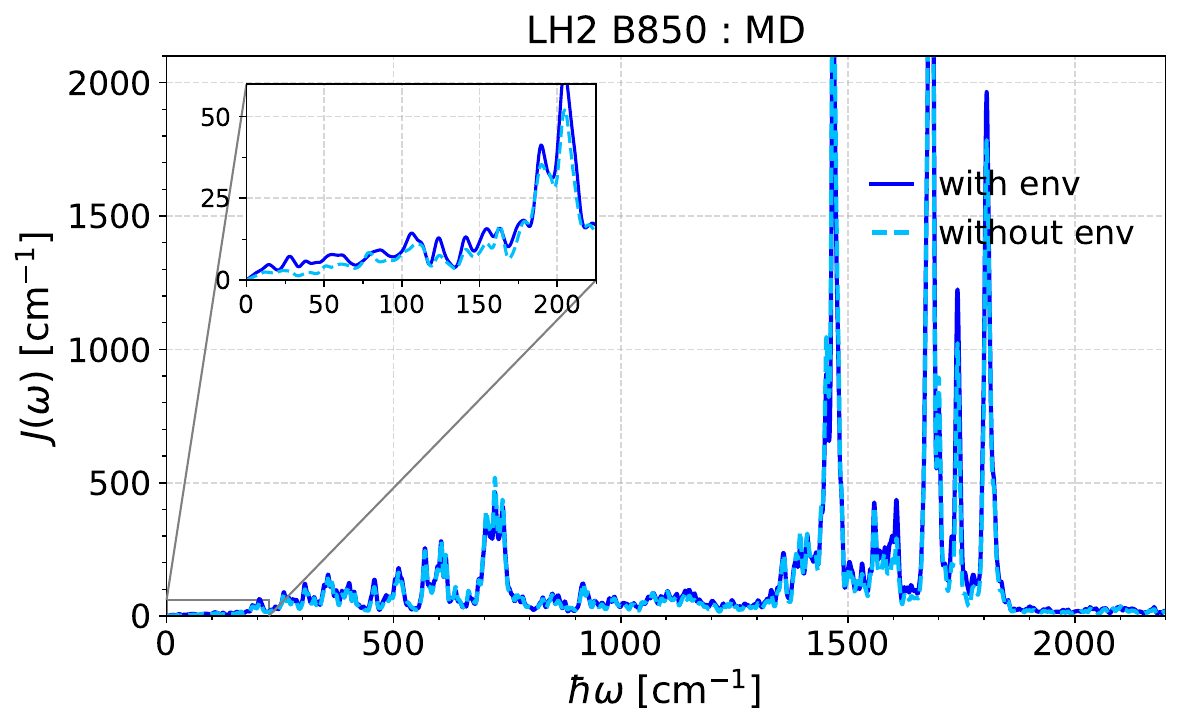}
        \caption{}
        \label{fig:B850_md}
    \end{subfigure}
    \caption{Spectral densities of the studied BChl pigment from the B800 (A, C) and the B850 (B, D) rings in the LH2 complex based on DFTB/MM MD (A, B) and classical MD (C, D) trajectories.}
    \label{fig:spd_lh2}
\end{figure*}

An example of the LH2 complex, containing the B800 and B850 rings, is shown in Fig.~\ref{fig:lh2_sys}. As described in the Methods section and highlighted in the figure, we considered one BChl pigment from each ring, i.e., a B800 and a B850 pigment. The ground-state dynamics of the LH2 complex were determined either using DFTB/MM or classical MD simulations, followed by TD-LC-DFTB calculations performed with and without the protein environment. Subsequently,   the spectral density was extracted and the influence of the environment assessed. The respective spectral densities are shown in Fig.~\ref{fig:spd_lh2}. As shown in the upper panels, the spectral density obtained from DFTB/MM trajectories exhibits low-frequency contributions in both the B800 and B850 rings. However, for the case of the B800 pigment, the influence of the protein environment is pronounced in the low-frequency region.
In contrast, for the pigment from the B850 ring, the inclusion of the protein environment has only a minimal effect, as the low-frequency region is largely unaffected by protein conformational dynamics. This indicates that the dominant contributions in the low-frequency region arise from slow vibrational motions of the BChl pigment itself, consistent with the gas-phase calculations discussed earlier. 
Moreover, the protein conformational dynamics seem to influence the high-frequency vibrational modes up to about 1100 cm$^{-1}$ in the case of the B800 ring. 
At the same time, differences in the frequency range around 1000 cm$^{-1}$ have been observed, however, already in earlier studies. The spectral densities based on DFTB/MM dynamics take longer to converge (see, e.g., Fig.~S8 in Ref.~\citenum{mait20a}). We are aware of this issue, but sampling issues are not the topic of the present study and seem to be most severe in the mid-frequency range and more prominent for BOMD simulations. 
A similar trend is observed for spectral densities based on the classical MD trajectories, as shown in the lower panels of Fig.~\ref{fig:spd_lh2}. However, although the low-frequency modes of BChl and the influence of protein dynamics are clearly visible for the B800 ring in classical MD using the CHARMM-compatible force field, their intensities are lower compared to those obtained from DFTB/MM trajectories. Again, this might also be a force field-dependent finding.  For the B850 ring, the effect of the protein environment remains minimal in the low-frequency region, also for the MD simulation. Only slow intramolecular modes of BChl pigment are contributing in this region, similar to the DFTB/MM MD results, but with lower intensity in the case of classical MD simulation. This 
finding is consistent with earlier simulations using classical MD simulations showing only small outer-sphere reorganization energies for the pigments in the B850 ring \cite{hoff25a}. Furthermore, it should be noted that removing the electrostatic environment only during the excited-state calculations does not completely eliminate the influence of the protein. The ground-state dynamics is still performed in the presence of the protein, which introduces additional contributions particularly in the low-frequency region compared to a purely gas-phase spectral density, as shown in Fig.~\ref{fig:dftb_and_normal_modes}. Additionally, we calculated the total reorganization energies for the B850 and B800 rings based on both the DFTB/MM MD and the classical MD trajectory using 
\begin{equation} 
\lambda_{m} = \int_{0}^{\infty} \frac{J_{m}(\omega)}{\omega} d\omega \label{eq:reorg}~.
\end{equation}
The results are given in Table~\ref{tab:reorg}.

\begin{table}[tb]
\centering
\caption{Reorganization energies $\lambda$ for different pigments obtained from DFTB/MM MD and classical MD simulations.}
\label{tab:reorg}
\vspace{15pt}
\scalebox{1.35}{
\begin{tabular}{|c|c|c|}
\hline
Pigment & DFTB/MM MD & Classical MD  \\
\hline
B800 & 762.19 cm$^{-1}$& 550.65 cm$^{-1}$\\
B850 & 352.85 cm$^{-1}$& 331.58 cm$^{-1}$\\
\hline
\end{tabular}}
\end{table}

As one can see in the Table~\ref{tab:reorg}, the reorganization energy of B800 is significantly larger than that of the B850 ring in both QM/MM and classical MD trajectories. Moreover, it is well known in the literature that the B850 ring exhibits stronger inter-pigment wave-function overlap compared to the B800 ring\cite{olbr10a, cupe16b, cupe18b, mall18a}. This provides a classic example of a transient delocalization event in the B850 ring relative to the B800 ring, where $\lambda_{m} \leq V_{mn}$. Such behavior has recently been reported by us in Ref.~\citenum{hoff25a}.
\begin{figure} [ht!]
\centering \includegraphics[width=0.75\textwidth]{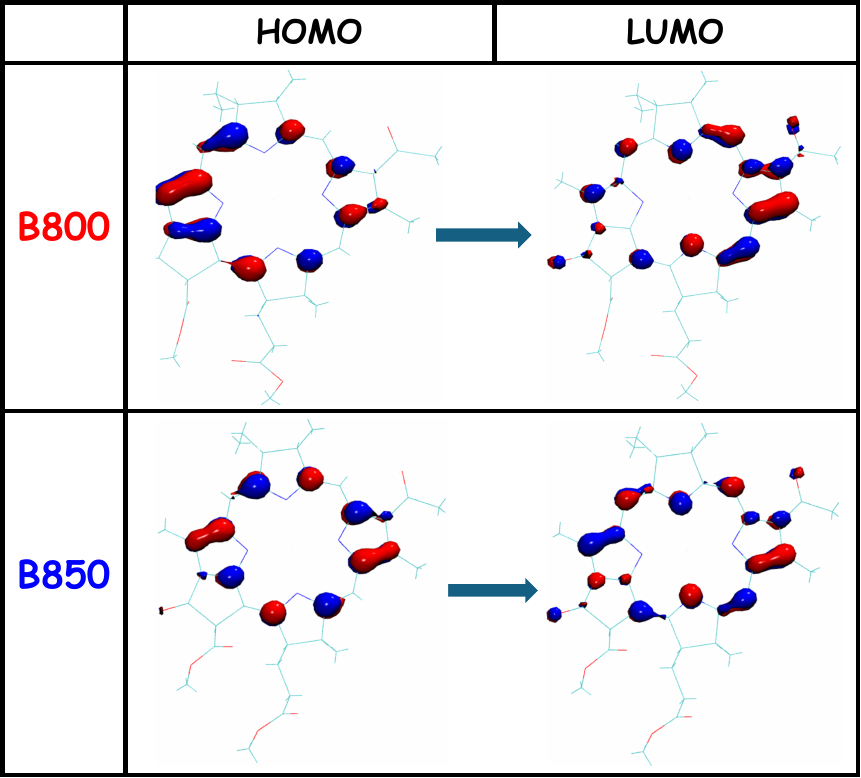}
\caption{\label{fig:homo_lumo_bcl} The HOMO and LUMO of the BChl pigments in the B800 and B850 rings were analyzed within the protein environment in a QM/MM setting. The molecular orbitals were visualized using an isosurface value of $\pm 0.05$ a.u.}
\end{figure}
Furthermore, to study the impact of large conformational dynamics of the surrounding environment on the spectral density of BChl pigments in the B800 and B850 rings, we examined the electrostatic effects of the environment on these pigments. Specifically, we analyzed how these effects influence the frontier orbitals, namely the highest occupied molecular orbital (HOMO) and the lowest unoccupied molecular orbital (LUMO). The pigments were first optimized within the DFTB/MM framework. Subsequently, time-dependent density functional theory (TD-DFT) calculations were performed on the optimized geometries using the $\omega$B97X/def2-TZVP level of theory within a QM/MM embedding. The resulting HOMOs and LUMOs are shown in Fig.~\ref{fig:homo_lumo_bcl}. 
As can be seen, the spatial distributions of the HOMO and LUMO orbitals of the BChl pigment in the B800 ring are localized on opposite sides of the molecule: while the HOMO orbital in Fig.~\ref{fig:homo_lumo_bcl} exhibits stronger localization on one side, the LUMO orbital shows a shifted localization toward the opposite side. In contrast, for the pigment in the B850 ring, the HOMO orbitals are more delocalized over the complete pigment, exhibiting a more symmetric distribution that is similar to that of the LUMO orbital. Moreover, a natural transition orbital (NTO) analysis further reveals that this HOMO-LUMO pair contributes roughly 85\% to the S$_1$ electronic excitation of the BChl pigment. Consequently, the S$_1$ energy gap is largely determined by this pair of orbitals, and its variations along the trajectory mainly stem from changes in these orbitals driven by environmental interactions. This asymmetric spatial localization of the frontier orbitals implies that electrostatic fluctuations of the protein matrix modulate the ground and excited states of the BChl pigment in the B800 ring differently, leading to pronounced excitation energy fluctuations and a strong contribution to the low-frequency features of the spectral density. In contrast, the frontier orbitals of the B850 pigment exhibit a largely symmetric spatial distribution, such that fluctuations in the protein electrostatics affect both states similarly and therefore have a negligible impact on the excitation energy gap.
%Furthermore, we calculated the absolute dipole moments of the ground and excited states for the pigments in the B800 and B850 rings as presented in Table~\ref{tab:dipole_moments}. The BChl pigments in the B800 ring exhibit larger dipole moments and a greater change between the ground and excited states compared to those in the B850 ring. This observation further supports the asymmetric and symmetric distributions of the HOMO and LUMO in B800 and B850 rings as discussed above.

To conclude the discussion of the character of the first excited states, we want to investigate a possible limitation in our methodology. Being centered around the Frenkel Hamiltonian Eq.\ref{eq:hamil}, which describes local excitations, dimer states such as charge-transfer (CT) states and their mixing into the local excitations are neglected. Experimental results\cite{beek97a, raet14a}, as well as recent theoretical studies, have however shown that the wavefunctions of neighboring B850 pigments overlap significantly and CT states are mixing into the local excitations \cite{nott18a,cupe22a}. Indeed, it has been shown that the inclusion of CT states into the system Hamiltonian is necessary to correctly describe the splitting and broadening of B800 and B850 bands. To assess the interaction between the neighboring pigments, we conducted DFT calculations on  monomer and dimer configurations of both B800 and B850 and extracted the dipole difference between the ground and first excited state (see Table~\ref{tab:dipole_moments}). In the case of B800 BChl, we find that the dipole moment change remains almost constant between the monomer and dimer calculation and agrees well with experimental value\cite{beek97a}. This suggests that both calculations describe the same local excitations and the CT character of this state seems to be very limited. This is different for the B850 BChl. Here, the calculation on the monomer yields a significantly lower dipole than the dimer calculation, which is also in reasonable agreement with the experimental result\cite{beek97a}. The strong impact of the inclusion of the dimer partner suggests that, in agreement with literature, the first excited state of the B850 band is shaped by the mixing with higher-lying CT states. From a theoretical point of view, local excitations of the monomers including CT components can be constructed in the Multistate Fragment-Excitation-Difference/Fragment-Charge-Difference approach \cite{nott18a,cupe18b}. Here, a large number of adiabatic excited states of the dimer are calculated and mapped onto a diabatic basis encoding local excitations and CT states. Having obtained the diabatic Hamiltonian, perturbation theory can be used to determine the involvement of the higher-lying CT states in the local excited states \cite{geme23a}. The application of this computationally demanding workflow to computations along a trajectory, that is for several thousand conformations, as necessary for the present spectral density calculations, is  beyond the scope of this study. It is, however,  most likely that inclusion of additional effects will increase the fluctuations and thus amplify the contributions in the low-frequency region.

%For the purpose of this study, highlighting the relevance of the low frequency region as a part of the intra-molecular contribution, we consider the Frenkel model to be sufficient. Higher-lying CT states, potentially mixed into the local excitations, are expected to only amplify the contributions in the low-frequency region. This is due to the strong coupling of the CT state's charge shift with the protein matrix. It is possible, that this would reduce the difference in the electrostatic effects of the protein observed between B800 and B850 here.

\begin{table}[tb]
\centering
\caption{\label{tab:dipole_moments} Ground (S$_0$) and excited-state (S$_1$) dipole moments $\mu$ and their absolute differences (in Debye) for BChl pigments in the B800 and B850 rings.} 
\vspace{15pt}
\scalebox{1.35}{
\begin{tabular}{|l|c|c|c|}
\hline
Pigment & $\mu(S_0)$ & $\mu(S_1)$  & |$\Delta \mu$|  \\
\hline
B800 monomer & 12.1406 & 8.3983 & \textbf{3.7422} \\
B800 dimer & 27.7255 & 23.6444 &  \textbf{4.0811} \\
\hline

B850 monomer & 3.0399  & 2.9021 &  \textbf{0.1377} \\
B850 dimer & 4.4247 & 1.9681 &  \textbf{2.4566} \\

\hline
\end{tabular}}
\end{table}

In addition, we calculated the root-mean-square deviation (RMSD) and site-energy fluctuations of the two pigments along the DFTB/MM MD trajectories (see Fig.~S4 in the SI). Both quantities are significantly larger for the B800 pigment than for the B850 pigment, indicating that the BChl pigment is more flexible in the binding pocket of the B800 ring than in that of the B850 ring. The pigment in the B850 ring is wedged in between the helices and other pigments with a fairly stable secondary structure. The BChl molecule in the B800 ring, on the other hand, is kept in place only by a smaller helix. Moreover, it is known that the surrounding protein environment of the BChl pigment in the B850 ring involves an axial, non-covalent, and rigid coordination with a histidine residue, while the pigment in the B800 ring coordinates electrostatically with a negatively charged aspartate residue, which is more dynamic in nature\cite{olbr10a,mall18a}.

%Furthermore, the B850 ring is tightly packed, whereas the pigments in the B800 ring are separated by a larger distance, which also makes them more flexible.
Thus, the combined effects of increased flexibility of the BChl pigment within the binding pocket and enhanced sensitivity of the frontier orbitals to the electrostatic environment lead to a stronger influence on the excitation energy fluctuation and resulting spectral density of the BChl pigment within the B800 ring. A similar effect was observed in a previous study for BChl pigments in the FMO complex, where strong coupling between the low-frequency vibrational modes of the BChl molecule and the protein conformational dynamics was reported based on classical MD simulations\cite{aght14a}. This effect is further confirmed in the present study below, but with conformations based on a DFTB/MM MD trajectory.

\subsection{Spectral Density of BChl Pigment in the FMO Complex}

\begin{figure} [ht!]
\centering \includegraphics[width=0.60\textwidth]{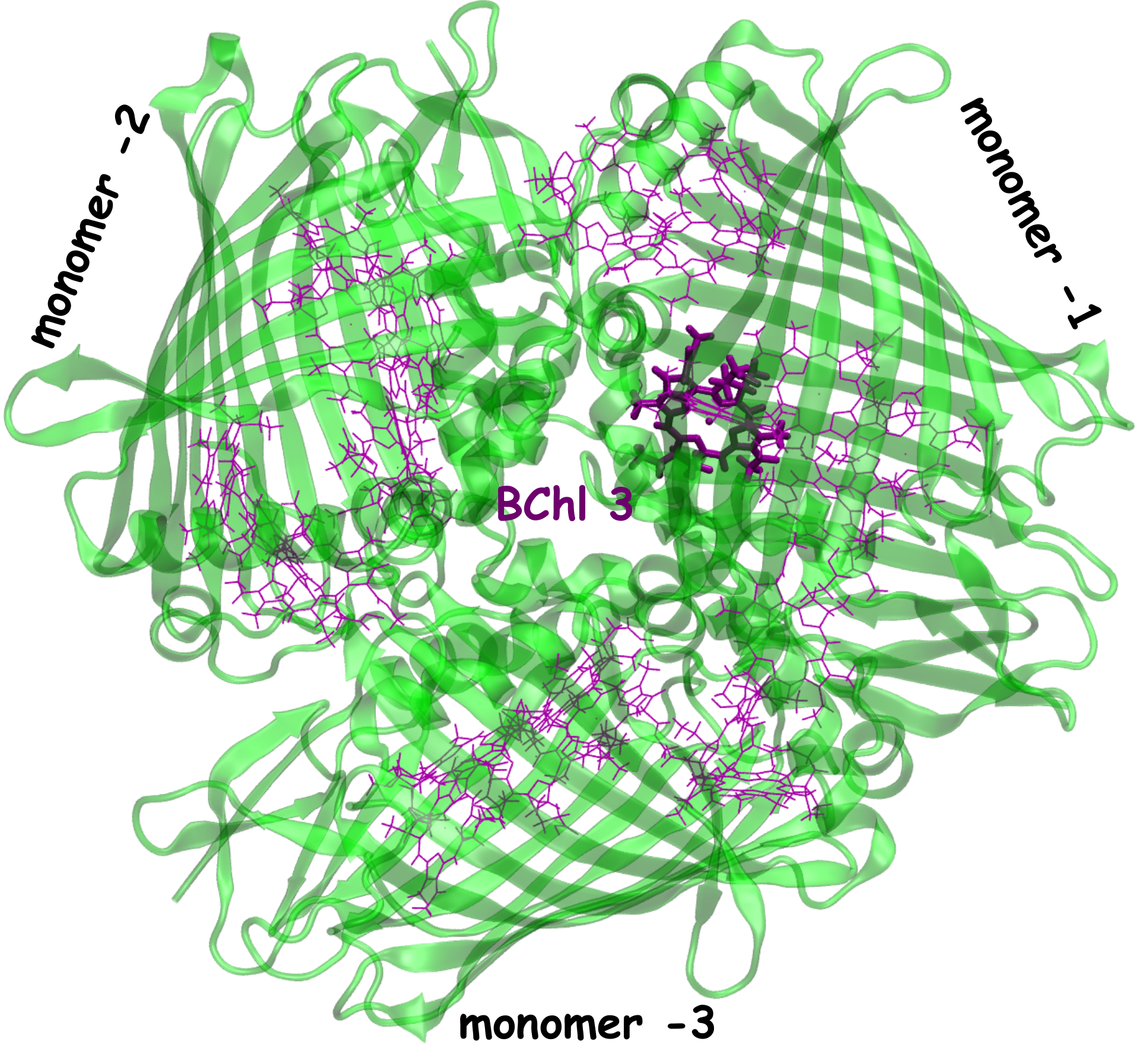}
\caption{\label{fig:fmo_sys} An example of the trimeric FMO complex from the green sulfur bacterium C. tepidum (PDB code: 3ENI) is shown, containing a total of 24 bacteriochlorophyll pigments (purple). In one monomeric unit, the BChl 3 pigment, considered as the QM region in this study, is highlighted.}
\end{figure}

\begin{figure*}[ht!]
    \centering

    % Left panel
    \begin{subfigure}[b]{0.50\textwidth}
        \centering
        \includegraphics[width=\textwidth]{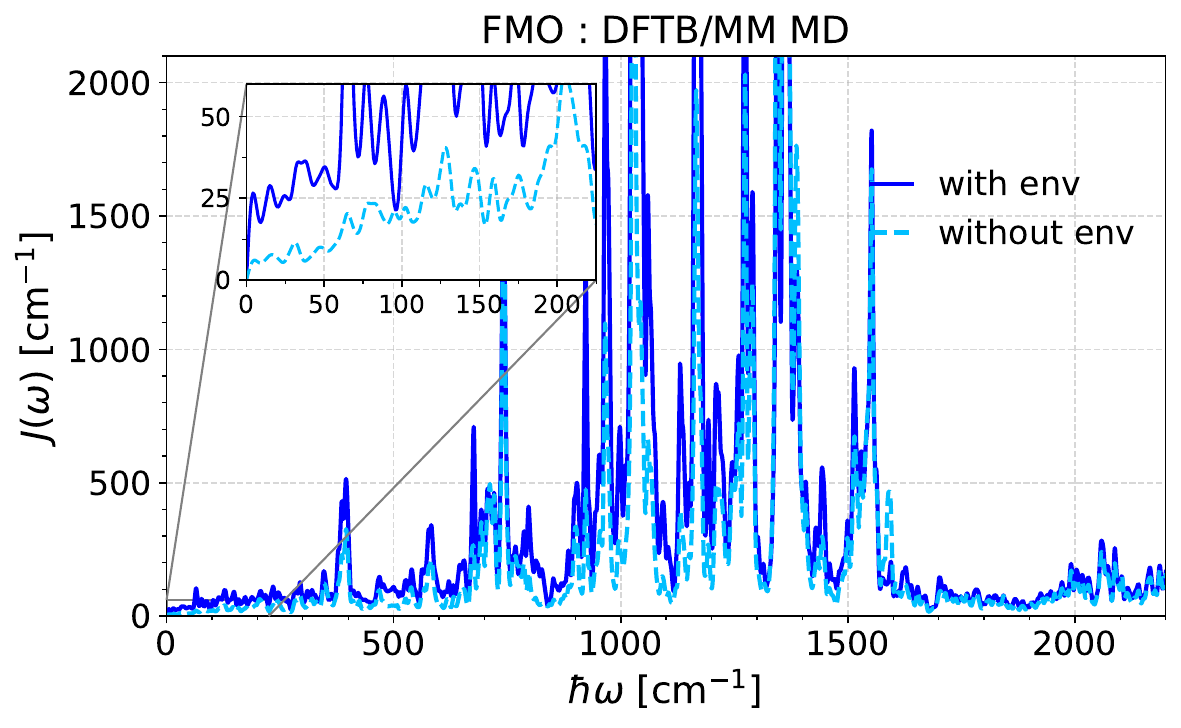}
        \caption{}
        \label{fig:FMO_dftb}
    \end{subfigure}
    \hfill
    % Right panel
    \begin{subfigure}[b]{0.49\textwidth}
        \centering
        \includegraphics[width=\textwidth]{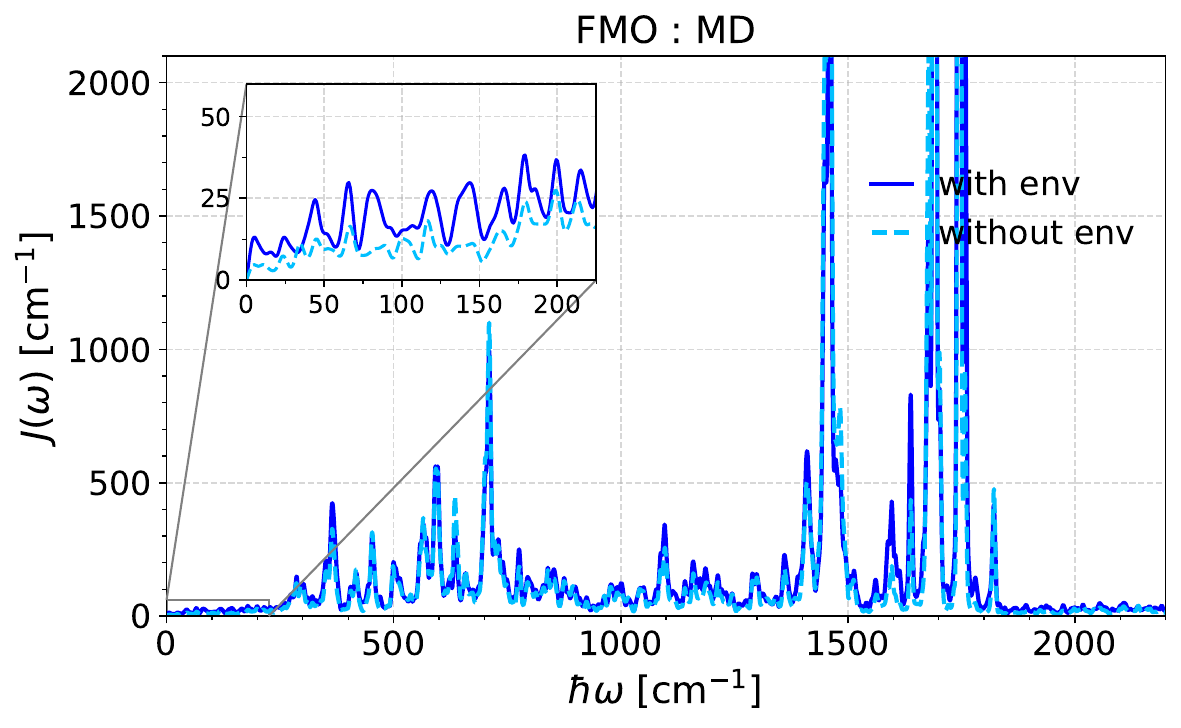}
        \caption{}
        \label{fig:FMO_md}
    \end{subfigure}

    \caption{Spectral densities of a BChl pigment in the FMO complex based on DFTB/MM MD (A) and classical MD (B) trajectories from our previous study\cite{mait20a}.}
    \label{fig:spd_fmo}
\end{figure*}

Similar to the LH2 complex, another extensively experimentally and computationally studied model LH system is the FMO complex of green sulfur bacteria\cite{enge07a,ishi09a,olbr11a,olbr11b, thyr18a,cao20a}. In this work, we consider the equilibrated FMO structure of \textit{C. tepidum} obtained from our previous study \cite{mait20a}. The FMO complex is a trimeric assembly, with each monomeric unit containing eight BChl pigments. A representative structure of the trimeric FMO complex is shown in Fig.~\ref{fig:fmo_sys}. Here, we consider BChl 3 as a representative example to examine the effect of protein conformational dynamics and to analyze the slow vibrational modes of the pigment. To this end, both DFTB/MM and classical MD trajectories were computed, followed by TD-LC-DFTB calculations with and without environmental point charges. The extracted spectral densities are shown in Fig.~\ref{fig:spd_fmo}.

As can be seen, the low-frequency regions of the spectral densities based on both DFTB/MM and classical MD trajectories are strongly influenced by the protein environment, as evidenced by the clear differences between the results with and without QM/MM coupling. In the case of the DFTB/MM MD trajectories, the intensities of the low-frequency peaks are stronger than those obtained from classical MD, which again might be dependent on the employed force field.  A similar trend was observed within the LH2 complex for excited state calculations with and without QM/MM coupling. Furthermore, the influence of the protein environment in the FMO complex is comparable to that observed for the pigments in the B800 ring of the LH2 complex, as shown in the previous section. Moreover, in a previous investigation using excitation energies based on Zerner’s intermediate neglect of differential orbital
method with spectroscopic parameters together with configuration interaction using single excitations (ZINDO/S-CIS) (commonly termed  ZINDO/S) calculations along classical MD trajectories, we showed a strong influence of the conformational dynamics of the protein on the BChl pigments in the FMO complex. Therefore, we repeated the same ZINDO/S calculations on the classical MD trajectories generated in our earlier study\cite{mait20a} for all eight BChl pigments in a monomeric unit and extracted the spectral densities both with and without environmental point charges. The corresponding comparison is shown in the Supporting Information (see Fig.~S5). All pigments exhibit a pronounced influence of the protein dynamics as can be seen when the point charges are turned off. Since ZINDO/S is known to overestimate excitation energy fluctuations within a protein matrix\cite{mait20a}, the resulting spectral densities display an enhanced intensity in the low-frequency region compared to the TD-LC-DFTB calculations shown in Fig.~\ref{fig:spd_fmo}. Nevertheless, even in the absence of environmental point charges, low-frequency vibrational features persist in calculations based on classical MD-based trajectories, indicating the presence of slow intramolecular and pigment-protein coupled motions of the BChl pigments, which are important to the excitation dynamics.

\section{\label{sec:conclusion} Conclusions and Outlook}
Spectral densities are a key quantity for studying exciton dynamics in light-harvesting complexes within the framework of an open quantum system. The low-frequency region of the spectral density, which plays a crucial role in the excitation dynamics and the optical properties of the pigment network, is commonly attributed to slow conformational motions of the protein environment. However, although bacteriochlorophyll molecules are often considered rigid in the context of spectral densities, we find in this study that low-frequency contributions can arise from slow intramolecular vibrational motions, primarily associated with distortion-related vibrations of the pigment itself. These slow modes are intrinsic features of the porphyrin ring, such as doming, ruffling, saddling, and waving, as observed in X-ray crystallographic measurements\cite{jent97a,shel98a,jent98a}. This observation has been verified using semi-empirical methods, density functional theory calculations, and classical molecular dynamics simulations. All approaches reveal the existence of low-frequency modes in the spectral density already in vacuum. However, in classical molecular dynamics using a commonly employed CHARMM-compatible force field, these modes appear with a significantly lower intensity compared to Born Oppenheimer molecular dynamics simulations based on DFTB or DFT methods. Moreover, we further analyzed the NMA results and found that the low-frequency modes are only partially captured. While this observation supports the presence of these modes, the harmonic approximation used in NMA fails to fully reproduce them compared to analyses based on site-energy fluctuations.

We further investigated the influence of the protein dynamics on the spectral density of BChl pigments in the B800 and B850 rings of the LH2 complex as well as in the FMO complex. Our results show that the protein matrix in the B800 ring is highly flexible and strongly modulates the low-frequency region of the spectral density, whereas its effect in the B850 ring is negligible. Further analysis indicates that the electron densities associated with the HOMO and LUMO orbitals of the BChl molecules in the B800 ring are spatially distinct. Moreover, the dipole moments of the ground and excited states differ significantly in the B800 ring. This leads to different electrostatic interactions
of the ground and excited states with fluctuating charges of the environment, i.e., the influence of the protein on the energy gap.
A similar effect can be observed for the BChl pigments in the FMO complex. These findings indicate that the low-frequency modes of the spectral density arise from a combination of slow intramolecular vibrations of the pigment and conformational dynamics of the surrounding protein matrix in the FMO complex and the B800 ring of the LH2 complex. Surprisingly, this is different for the pigments of the B850 ring. There, the electron densities of the HOMO and LUMO orbitals of the BChl molecules are spatially quite similar. Moreover, the dipole moments of the ground and excited states are also comparable, implying that the ground- and excited-state energies are influenced in a similar manner by the protein environment. Thus, the surrounding point charges have almost no influence on the energy gap and thus on the spectral density. A similar effect was found earlier for bilin molecules in the PE545 light-harvesting complex  \cite{aght14a}.

Overall, this study highlights an important feature of apparently rigid molecules such as BChl in bacterial LH antenna complexes, and more generally of porphyrin-containing pigment molecules. It is often assumed that the low-frequency modes in the spectral density of BChl or Chl molecules originate mainly from conformational dynamics of the surrounding environment. However, our results show that these modes arise from a combination of both protein conformational dynamics and slow intramolecular vibrations of the pigment molecules themselves. Thus, determining the low-frequency part of the spectral density based on coupling to the environment only might be problematic as is done, e.g., in the charge density coupling (CDC) approach \cite{lee17b,lee16c}. An accurate determination of the low-frequency modes is also tricky in a simple normal mode analysis \cite{reng12b,reng13a}. 
Instead, a full description of the structured spectral density requires accounting for excitation energy fluctuations, where the ground-state dynamics captures all internal vibrations of the pigment molecules, including the low-frequency ones, and together with a QM/MM treatment during both the ground state dynamics and the excited energy calculations. Only by combining these treatments, one can ensure that all effects which contribute to the spectral density are taken care of.

In summary, this study has two key findings. The first one is that internal low-frequency modes of BChl molecules do contribute to spectral densities. Thus, the low-frequency part of the spectral densities is not determined only by the coupling to the environment. The second finding is that in certain environments, the ground and excited states of pigment molecules can be influenced in the same way by an electrostatic environment, i.e., the energy gap is not influenced by protein fluctuations. This was known for bilin molecules in the PE545 complex\cite{aght14a}, but the same can be true for BChl molecules, e.g.,  in the B850 ring of LH2 complexes.   

\section*{Supplementary Material}
See the Supplementary Material for further details on Huang-Rhys factors from gas-phase spectral density and normal-mode analysis, comparison with experimental measurements, additional gas-phase DFT spectral densities, and protein effects from ZINDO/S calculations along a classical MD trajectory for the FMO complex. RMSD and density of states plots for the BChl pigments in the B800 and B850 rings of the LH2 complex are also provided.

\section*{Acknowledgments} 
Financial support by the Deutsche Forschungsgemeinschaft (DFG) through grants  KL-1299/24-1 and KL-1299/25-2 is gratefully acknowledged. Furthermore, the calculations and simulations were performed on a compute cluster funded through the project INST 676/7-1 FUGG.

\section*{AUTHOR DECLARATIONS} 
\section*{Conflict of Interest} 
The authors have no conflicts to disclose.

\section*{Data availability} 
The data that support the findings of this study are available from the corresponding author upon reasonable request.

\section*{Code availability} 
The authors declare that the present research has been primarily produced using publicly available software, as detailed in Section \textbf{\ref{sec:method}}. Furthermore, the spectral density was calculated using our custom code, which is available at \href{https://github.com/CPBPG/TorchNISE}{https://github.com/CPBPG/TorchNISE}. Additionally, the spectral density data are available on Zenodo at 
\href{https://doi.org/10.5281/zenodo.19094107}{https://doi.org/10.5281/zenodo.19094107}.

\bibliography{ukleine}%,ukleine_add}% Produces the bibliography via BibTeX.
\end{document}

\section{Huang-Rhys Factors for Gas-Phase Calculations}
\begin{figure} [ht!]
\centering \includegraphics[width=0.80\textwidth]{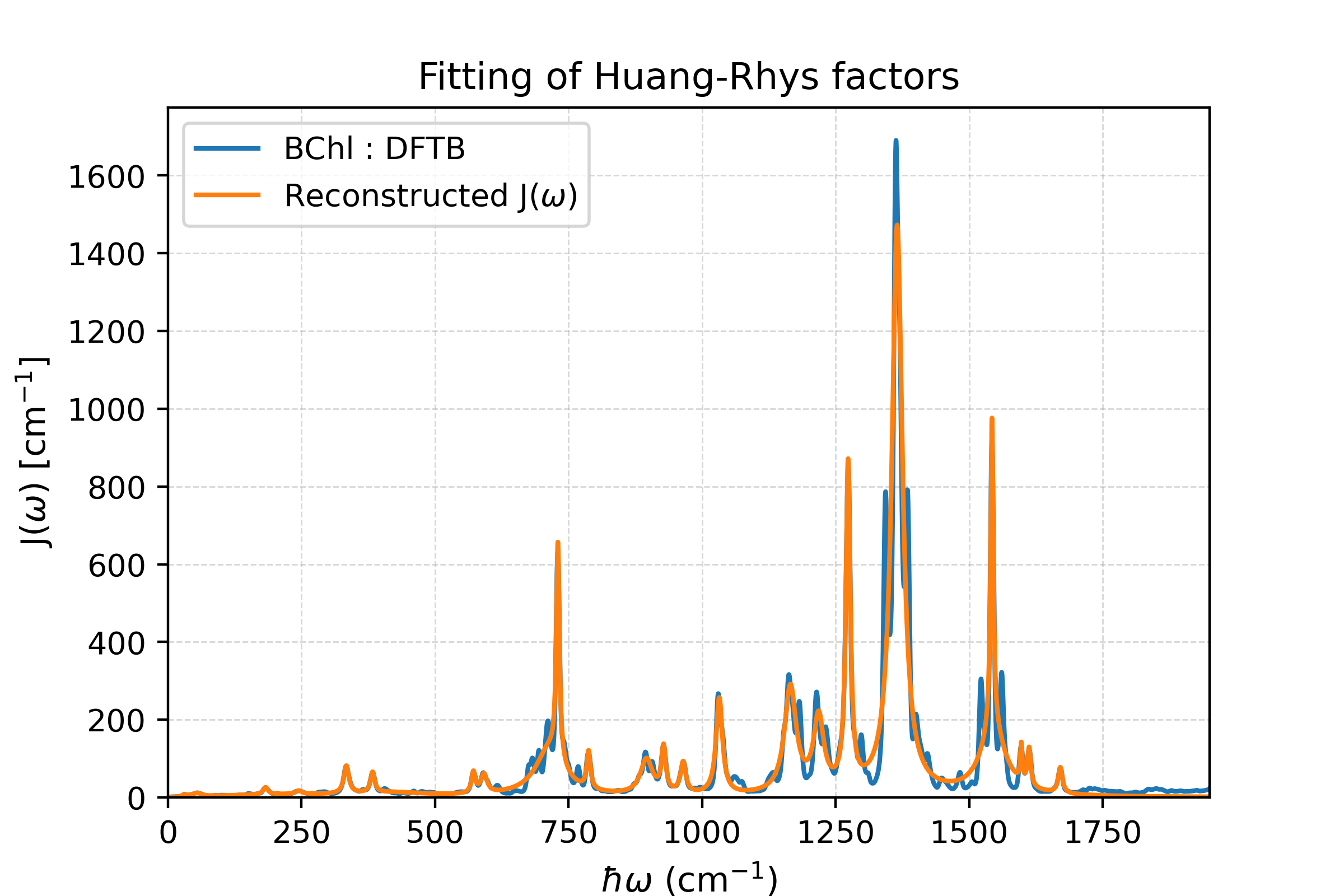}
\caption{\label{fig:dftb_fitted_hr} For the derivation of HR factors of the DFTB/3OB-f based spectral density, 26 peaks were manually identified and the positions, Huang-Rhys factors, and peak widths were optimized in reference to the DFTB-based spectral density. A comparison between the original and the reconstructed spectral density is shown.}
\end{figure}

In this section, the Huang-Rhys (HR) factors obtained from normal mode analysis (NMA) and the approximate HR factors derived from gas-phase DFTB/3OB-f calculations are presented. Due to the broadened spectral density produced by the DFTB-based approach, only an approximate assignment of vibrational modes and their corresponding HR factors is feasible. To address this, 26 peak positions were manually selected and assigned uniform HR factors. Subsequently, the least-squares fitting routine of \textit{SciPy} was applied to the reconstructed spectral density, based on Eq.~\ref{eq:construct_vib_sd}, in order to optimize the peak positions, HR factors, and peak widths with respect to the original spectral density, as defined by
\begin{equation}
\min_{\omega_k,\lambda_k,\gamma_k} \left| J_{\text{DFTB}}(\omega) - J^{\text{exp}}_{\text{vib}}(\omega_k,\lambda_k,\gamma_k) \right|^2~.
\label{eq:re_reorg}
\end{equation}
To allow for a stable fitting, the range of $\omega $ was limited from 0 to 1800 $\text{cm}^{-1}$. The spectral density reconstructed from the final set of modes, HR factors, and peak widths is shown in Fig. \ref{fig:dftb_fitted_hr}. The total reorganization energy is reproduced within 1.5\% of the original value. As the reconstructed spectral density was derived through a multi-dimensional fitting procedure, the mode positions and Huang-Rhys factors have to be understood as a minimal, sufficient description of the target spectral density.
The mode frequencies and corresponding HR factors obtained from the DFTB simulations are listed in Table~\ref{tab:DFTB_hr}, while those derived from the NMA are presented in Tables~\ref{tab:NMA_hr_lowF} and \ref{tab:NMA_hr}, respectively.

\newpage
\begin{center}
\begin{longtable}{|c|c|c|c|}
\caption{\label{tab:DFTB_hr} Huang-Rhys factors for the gas-phase spectral density obtained at the DFTB/3OB-f level of theory. The HR factors and corresponding vibrational mode positions were optimized through least-squares fitting to reproduce the original spectral density.} \\
\hline
Mode frequency (cm$^{-1}$) & Huang-Rhys factor & Mode frequency (cm$^{-1}$) & Huang-Rhys factor\\
\hline

31.61 &0.0951 &895.92 &0.0049 \\
55.15 &0.0984& 927.84 &0.0021\\
182.83 &0.0098& 964.52 &0.0016\\
245.23 &0.0034 &1032.11 &0.0052\\
261.61 &0.0433 &1165.25 &0.0084\\
333.97 &0.0114 &1218.15 &0.0046\\
383.05 &0.0052 &1273.26 &0.0079\\
423.20 &0.0143& 1365.11 &0.0295\\
571.79 &0.0022& 1542.60 &0.0031\\
591.52 &0.0031& 1544.69 &0.0061\\
714.80 &0.0236 &1597.44 &0.0003\\
729.92 &0.0113& 1612.17 &0.0004\\
787.38 &0.0020& 1670.51 &0.0004\\

\hline
\end{longtable}
\end{center}

\begin{center}
\begin{longtable}{|c|c|c|c|}
\caption{\label{tab:NMA_hr_lowF} Mode resolved Huang-Rhys factors from the normal modes analysis at CAM-B3LYP/def2-TZVP level of theory  for $\omega_k<230\:\text{cm}^{-1}$. All modes with an HR factor above 0.001 are shown.} \\
\hline
Mode frequency (cm$^{-1}$) & Huang-Rhys factor & Mode frequency (cm$^{-1}$) & Huang-Rhys factor\\
\hline
17.94 & 0.004 & 123.72 & 0.003 \\
18.95 & 0.003 & 130.50 & 0.002 \\
27.04 & 0.003 & 147.28 & 0.001 \\
29.72 & 0.003 & 159.50 &  0.001 \\
33.36 & 0.001 & 166.92 & 0.003 \\
37.36 & 0.001 & 175.41 & 0.001 \\
43.86 & 0.001 & 178.71 & 0.008 \\
47.80 & 0.001 & 193.83 & 0.002 \\
72.67 & 0.001 & 198.71 & 0.011 \\
79.58 & 0.003 & 203.51 & 0.001 \\
92.74 & 0.002 & 223.61 & 0.003 \\
95.36 & 0.001 & 227.97 & 0.004 \\
106.64 & 0.002 & & \\
\hline
\end{longtable}
\end{center}

\begin{center}
\begin{longtable}{|c|c|c|c|}
\caption{\label{tab:NMA_hr} Same as Table~\ref{tab:NMA_hr_lowF} but for $\omega_k> 230\:\text{cm}^{-1}$. All modes with  $\omega_k< 2000\:\text{cm}^{-1}$ and an HR factor above 0.001 are shown. Extreme high-frequency C-H vibrations in the 3000-3300 cm$^{-1}$ range are not provided.} \\
\hline
Mode frequency (cm$^{-1}$) & Huang-Rhys factor & Mode frequency (cm$^{-1}$) & Huang-Rhys factor\\
\hline
238.80&   0.004&     816.02&    0.005\\        
243.55&   0.009&     853.36&    0.002\\         
244.50&   0.009&     864.81&    0.003\\           
251.04&   0.003&     881.55&    0.002\\           
257.01&   0.008&     890.39&    0.013\\               
258.10&   0.003&     910.92&    0.001\\             
266.69&   0.001&     916.61&    0.014\\              
284.77&   0.001&     950.67&    0.002\\    
290.78&   0.002&     980.07&    0.006\\     
297.70&   0.001&     996.02&    0.002\\    
302.45&   0.002&     1000.65&   0.001\\        
317.33&   0.001&     1012.95&   0.002\\       
320.60&   0.006&     1021.10&   0.002\\       
1045.39&  0.001&     1036.61&   0.004\\   
328.05&   0.033&     1045.39&   0.001\\     
337.75&   0.002&     1050.15&   0.002\\    
356.15&   0.030&     1063.79&   0.003\\     
375.24&   0.008&     1102.80&   0.005\\          
380.50&   0.002&     1128.24&   0.001\\     
387.32&   0.007&     1136.76&   0.003\\     
411.76&   0.009&     1151.38&   0.004\\
428.27&   0.005&     1204.72&   0.005\\     
439.44&   0.019&     1214.41&   0.011\\        
449.59&   0.013&     1220.47&   0.002\\   
477.53&   0.002&     1228.59&   0.001\\   
552.32&   0.001&     1230.89&   0.001\\         
579.93&   0.007&     1241.17&   0.004\\           
596.01&   0.003&     1249.12&   0.003\\   
603.13&   0.001&     1287.02&   0.005\\     
604.78&   0.005&     1292.78&   0.001\\         
611.28&   0.002&     1304.49&   0.003\\            
638.22&   0.002&     1321.54&   0.001\\        
647.00&   0.011&     1331.98&   0.004\\          
669.82&   0.020&     1348.41&   0.003\\         
702.40&   0.002&     1373.03&   0.001\\      
711.02&   0.019&     1388.50&   0.001\\      
723.50&   0.003&     1397.68&   0.001\\       
739.86&   0.009&     1416.05&   0.001\\              
747.08&   0.004&     1427.80&   0.001\\           
750.17&   0.002&     1435.23&   0.001\\            
754.73&   0.001&     1453.77&   0.017  \\         
758.71&   0.031&     1566.21&   0.001\\             
771.80&   0.023&     1592.39&   0.001\\     
777.65&   0.006&     1606.35&   0.001\\       
788.45&   0.001&     1628.78&   0.001\\            
793.32&   0.002&     1696.36&   0.004\\     
803.46&   0.004&     1772.44&   0.001\\
815.35&   0.003&            & \\
\hline
\end{longtable}
\end{center}

\section{Experimental Spectral Density of BChl Molecule}
\begin{figure} [ht!]
\centering \includegraphics[width=\textwidth]{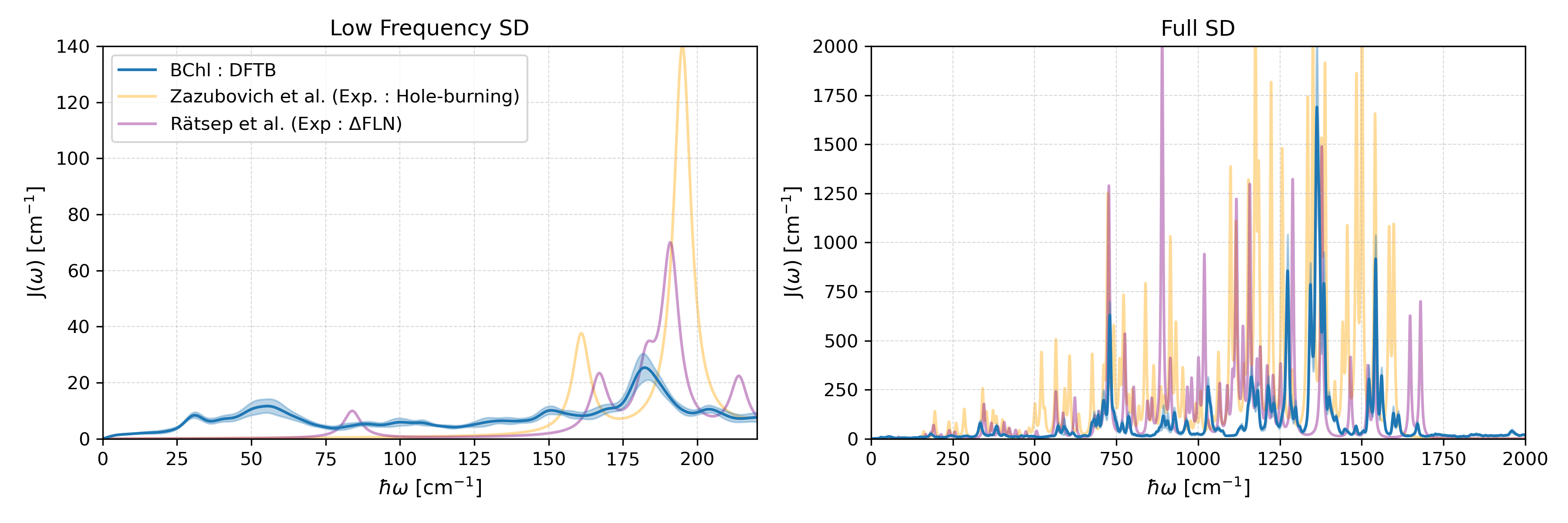}
\caption{\label{fig:dftb_exp_comparison} Comparison of the spectral density calculated in gas phase at the DFTB/3OB-f level with available experimental results obtained from hole-burning and $\Delta$FLN measurements. To this end, the spectral densities were constructed using experimentally determined mode-resolved Huang–Rhys factors according to Eq.~\ref{eq:construct_vib_sd}.}
\end{figure}

To assess the quality of the DFTB-based gas-phase spectral density, we compare it with available experimental data. However, since the experiments provide mode-resolved Huang-Rhys factors, these must first be converted into a structured spectral density for comparison. Here, we employ a Lorentzian expression considered in previous studies \cite{rosn15a,mait20a} to construct the structured spectral density
\begin{equation}
\label{eq:construct_vib_sd}
    J_{vib}^{exp}(\omega) = \frac{2\hbar}{\pi}\sum_k s_k \omega_k^3\frac{\omega \gamma_k}{(\omega_k^2-\omega^2)^2+\omega^2\gamma_k^2}~.
\end{equation}
In this expression, $s_k$, $\omega_k$, and $\gamma_k$ denote the mode-resolved Huang--Rhys factors, the normal-mode frequencies, and the mode broadening, respectively. The mode-resolved reorganization energy $\lambda_k$, used in the reconstruction of the spectral density in Eq.~\ref{eq:re_reorg}, is defined as the product of $s_k$ and $\omega_k$.
For the construction of the experimental spectral densities, we have considered a constant broadening of $\hbar \gamma = 7\,\text{cm}^{-1}$. The HR parameters were extracted from hole-burning experiments conducted by Zazubovich et al.~\cite{zazu02} and from the $\Delta$FLN experiments by Rätsep et al.~\cite{raet11a}. As illustrated in Fig.~\ref{fig:dftb_exp_comparison}, the simulated and experimental results show a satisfactory agreement in both peak positions and intensities for the hole-burning and $\Delta$FLN measurements. Nonetheless, the experiment resolves a larger number of high-frequency modes than the DFTB simulation, which can be attributed to sampling limitations in the BOMD simulation, as discussed in the main text.
Moreover, both experiments reveal pronounced low-frequency modes in the range of $150$--$250\,\text{cm}^{-1}$, suggesting the presence of slow vibrational motions. An additional peak around $75$-$100\,\text{cm}^{-1}$ is observed in the hole-burning experiment but is absent in the $\Delta$FLN measurements, indicating potential limitations of different experimental techniques in detecting all vibrational modes.

\section{Gas-Phase Spectral Density at DFT Level}
\begin{figure} [ht!]
\centering \includegraphics[width=0.75\textwidth]{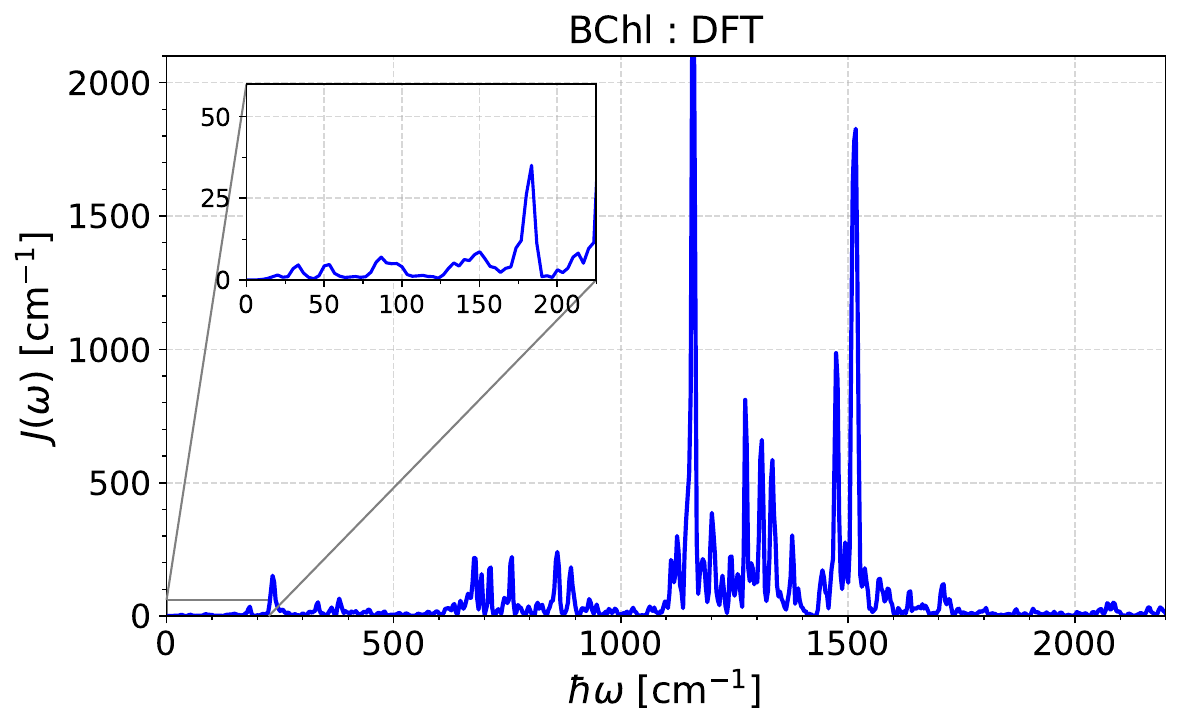}
\caption{\label{fig:dft_spd} Gas-phase spectral density of the BChl molecule calculated at BLYP/6-31G(d,p) level based on NVE ensembles.}
\end{figure}

As discussed in the main text, we further extracted the spectral density from a DFT-based simulation performed at the BLYP/6-31G(d,p) level of theory in the NVE ensemble, i.e., without a thermostat. Due to the high computational cost, only a single trajectory of 10~ps was generated. Further details are provided in the Computational Methods section of the main text. Figure~\ref{fig:dft_spd} shows the spectral density obtained from the DFT simulation. The result exhibits non-zero contributions in the low-frequency region, consistent with the DFTB simulations discussed in the main text. In addition, the high-frequency features are in good agreement with the DFTB results, as the 3OB-f parameter set is known to accurately reproduce C=C, C=O, and C=N vibrational modes compared to the BLYP functional. Overall, both approaches consistently capture the low-frequency modes associated with slow intramolecular vibrations of the relatively rigid BChl molecules. For the low frequency region ($\omega_k < 235\:\text{cm}^{-1} $) a sum of Huang-Rhys factors of 0.1315 was determined using the fitting procedure shown above.

\section{Fluctuations of BChl Pigment in B800 and B850 Rings}
\begin{figure*}[ht!]
    \centering

    % First row
    \begin{subfigure}[b]{0.50\textwidth}
        \centering
        \includegraphics[width=\textwidth]{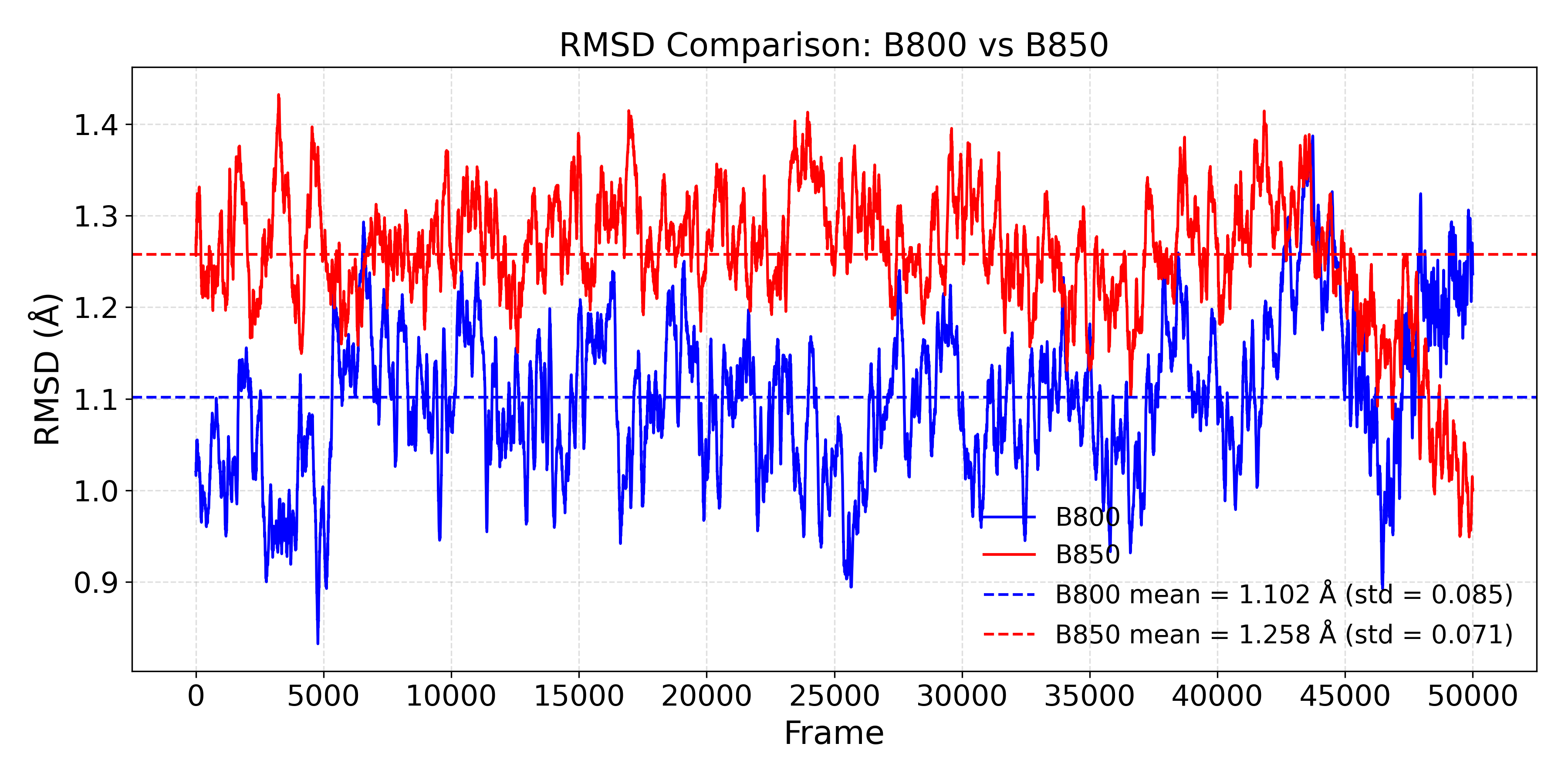}
        \caption{}
        \label{fig:rmsd}
    \end{subfigure}
    \hfill
    \begin{subfigure}[b]{0.49\textwidth}
        \centering
        \includegraphics[width=\textwidth]{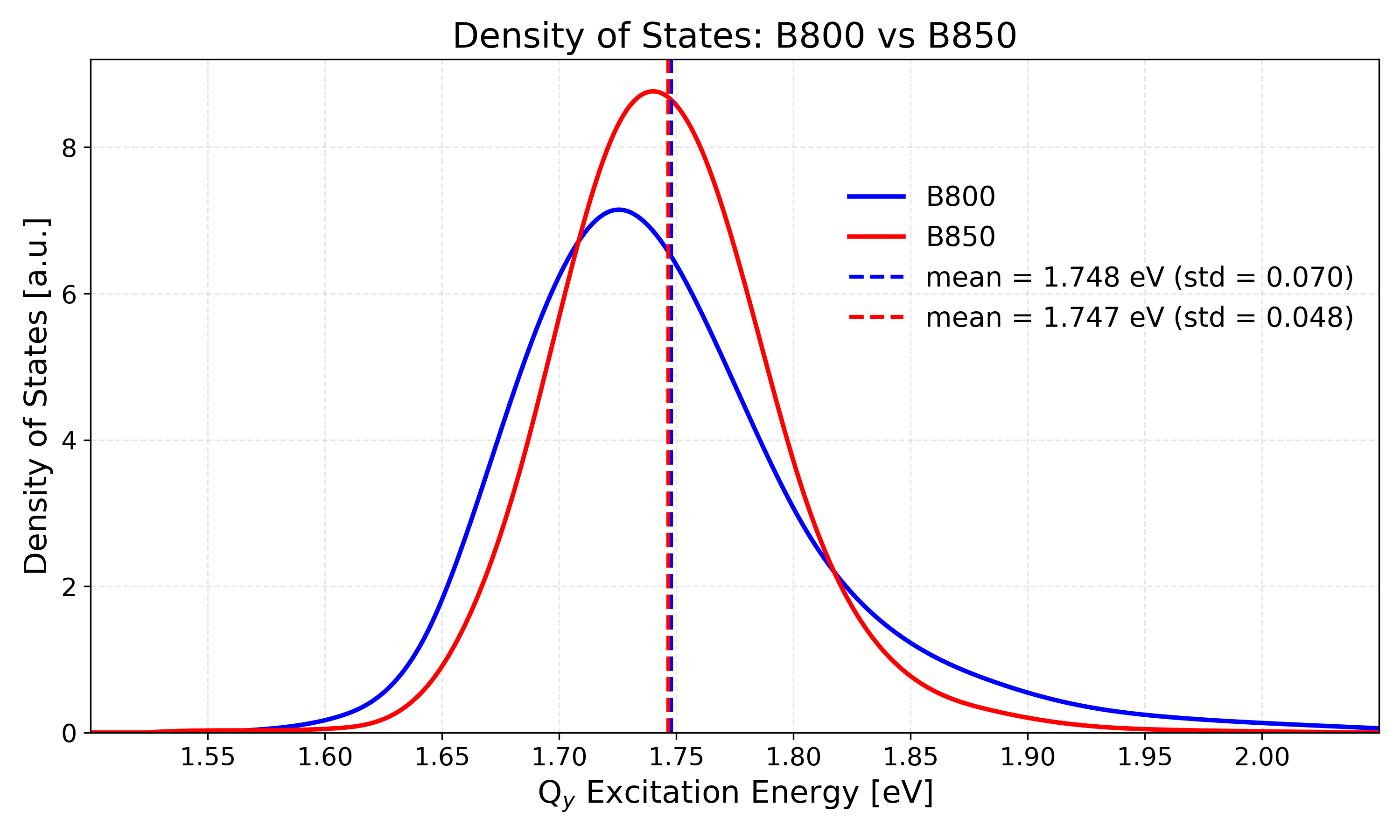}
        \caption{}
        \label{fig:dos}
    \end{subfigure}

    \caption{RMSD and the distribution of excitation energies of BChl pigments in the B800 and B850 rings of the LH2 complex are shown in panels A and B. The corresponding average values, together with the standard deviation, are also provided.}
    \label{fig:rmsd_dos}
\end{figure*}

We extracted the RMSD and the distribution of excitation energy fluctuations, referred to as the density of states (DOS), for the BChl pigments in the B800 and B850 rings, as shown in Fig.~\ref{fig:rmsd_dos}. Both the RMSD and the density of states exhibit larger fluctuations for the BChl pigment in the B800 ring than for that in the B850 ring. This structural flexibility of the BChl pigment within the protein binding pocket, induced by the surrounding protein matrix, together with the susceptibility of the frontier orbitals to the electrostatic environment, as discussed in the main text, enhances the excitation energy gap fluctuations and thereby increases the intensity of the spectral density, particularly in the low-frequency region.

\section{Impact of Protein Dynamics on the Spectral Density}
\begin{figure} [ht!]
\centering \includegraphics[width=\textwidth]{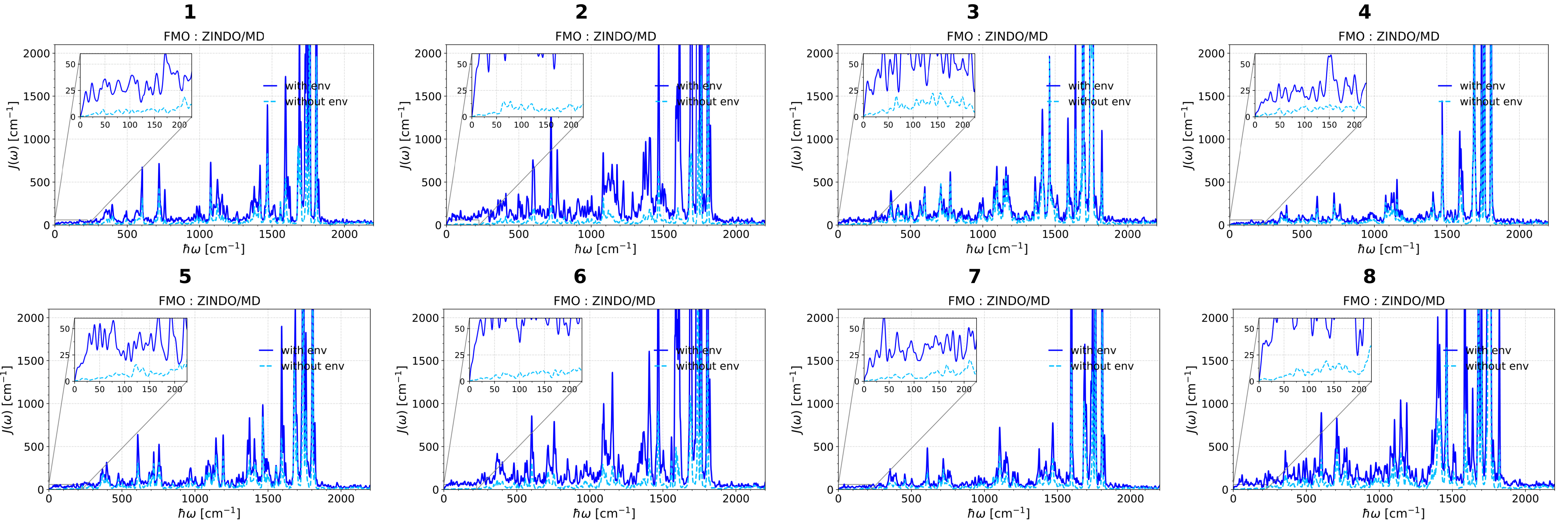}
\caption{\label{fig:fmo_spd} Spectral density of an individual pigment molecule within a monomeric FMO complex unit of the trimer, obtained from classical molecular dynamics trajectories followed by ZINDO/S calculations. The solid line represents the spectral density including the environment, while the dashed line corresponds to the calculation without the environment.}
\end{figure}

As discussed in the main text, we repeated the spectral density calculations for a well-studied model system, the FMO complex, for which low-frequency modes in the spectral density are known to originate from protein conformational dynamics. We employed methodologies previously used by our group\cite{damj02a,olbr10a,olbr11b} and by several others\cite{gao13a,vian14a,wang15a,zhen16a}, which are classical MD simulations followed by excitation-energy calculations using the semi-empirical ZINDO/S method on antenna complexes in a QM/MM framework. The extracted spectral densities for all eight BChl pigments within a single monomer of the trimer, obtained with and without QM/MM coupling, are presented in Fig.~\ref{fig:fmo_spd}.

This figure clearly illustrates that the low-frequency modes primarily originate from protein conformational dynamics when the QM/MM coupling is turned off in the excited-state calculations. This observation was also reported in our earlier study for the same complex using the same multiscale strategy \cite{aght14a}. However, a closer inspection shows that these low-frequency modes do not vanish completely in the absence of QM/MM coupling; instead, their intensities are significantly reduced. As discussed in the main text, this behavior arises because the classical MD simulations based on a CHARMM-compatible force field for the BChl pigment do not adequately capture the intensity of the slow intramolecular modes of the pigment. In contrast, our previous study indicates that an AMBER-compatible force field might yield more intense peaks for these slow intramolecular modes of the BChl pigment\cite{chan15a}. Nevertheless, it is clear that the low-frequency modes of the FMO complex receive contributions from protein conformational dynamics as well as from pigment intrinsic modes, similar to the flexible B800 ring in the LH2 complex. This contrasts with the rigid B850 ring, where the low-frequency modes are dominated primarily by pigment intrinsic contributions, as discussed in the main text.

\bibliography{ukleine}%,ukleine_add}